\documentclass[twocolumn,showpacs,preprintnumbers,amsmath,amssymb,superscriptaddress,bibnotes,floatfix]{revtex4-1}

\usepackage{dcolumn}% Align table columns on decimal point
\usepackage{bm}% bold math
\usepackage[dvipdfmx,usenames,dvipsnames]{color}
\usepackage[dvipdfmx]{graphicx}

\newcommand{\LV}{\mathrm{LV}}

\newcommand{\LVin}{{\rm LV}^{\rm in}}
\newcommand{\LVout}{{\rm LV}^{\rm out}}
\newcommand{\CVin}{{\rm CV}^{\rm in}}
\newcommand{\CVout}{{\rm CV}^{\rm out}}
\newcommand*{\ex}{\mathrm{ex}}
\newcommand*{\self}{\mathrm{self}}

\begin{document}
\preprint{APS/123-QED}
\title{Input-output relationship in social communications characterized by spike train analysis}

\author{Takaaki Aoki}
\email{takaaki.aoki.work@gmail.com}
\affiliation{Faculty of Education, Kagawa University, Takamatsu 760-8521, Japan}
\author{Taro Takaguchi$^\dagger$}
\affiliation{National Institute of Informatics, 2-1-2 Hitotsubashi, Chiyoda-ku, Tokyo 101-8430, Japan}
\affiliation{JST, ERATO, Kawarabayashi Large Graph Project, 2-1-2 Hitotsubashi, Chiyoda-ku, Tokyo 101-8430, Japan}
\altaffiliation[Present address: ]{National Institute of Information and Communications Technology, 4-3-1 Nukui-Kitamachi, Koganei, Tokyo 184-8795, Japan}
\author{Ryota Kobayashi}
\affiliation{National Institute of Informatics, 2-1-2 Hitotsubashi, Chiyoda-ku, Tokyo 101-8430, Japan}
\affiliation{Department of Informatics, Graduate University for Advanced Studies (Sokendai), 2-1-2 Hitotsubashi, Chiyoda-ku, Tokyo 101-8430, Japan}
\author{Renaud Lambiotte}
\affiliation{Department of Mathematics and naXys, University of Namur, 8 Rempart de la Vierge, Namur B-5000, Belgium}
\date{\today}

\begin{abstract}
We study the dynamical properties of human communication through different channels, i.e., short messages, phone calls, and emails, adopting techniques from neuronal spike train analysis in order to characterize the temporal fluctuations of successive inter-event times. 
We first measure the so-called local variation (LV) of incoming and outgoing event sequences of users, and find that these in- and out- LV values are positively correlated for short messages, and uncorrelated for phone calls and emails. 
Second, we analyze the response-time distribution after receiving a message to focus on the input-output relationship in each of these channels.
We find that the time scales and amplitudes of response are different between the three channels.
To understand the impacts of the response-time distribution on the correlations between the LV values, we develop a point process model whose activity rate is modulated by incoming and outgoing events.
Numerical simulations of the model indicate that a quick response to incoming events and a refractory effect after outgoing events are key factors to reproduce the positive LV correlations. 
%\old{Finally, we also find that the LV value is mostly uncorrelated with conventional centrality measures of nodes in the aggregate network, suggesting that this type of analysis reveals a new dimension of social networks, associated to their temporal properties.} 
\end{abstract}
\pacs{02.50.-r,89.75.Fb, 89.75.Hc,89.65.-s}
\maketitle

%%%%%%%%%%%%%%%%
%%%%   Introduction
%%%%%%%%%%%%%%%%
\section{Introduction}\label{sec:intro}
%%% Background (change of perspective from adjacency matrix to an time sequence)
The study of social systems from a network perspective has a long tradition in the social sciences, i.e., social network analysis~\cite{wasserman}, and has played a central role in the recent advent of computational social science~\cite{Lazer2009}.
Focusing on the structure of social relationships has revealed generic properties of social networks, including their strongly heterogeneous connectivity and modular structures~\cite{Albert:2002p869,Newman:2003p839}.
Structural properties of social networks also exhibit considerable impacts on different types of dynamical processes on the networks, such as epidemic and information spreading~\cite{Barrat2008,Pastor-Satorras2015}.

As accessible datasets of human behavior become increasingly rich, various approaches have been employed to improve the network modeling in order to uncover hidden aspects of human dynamics.
%\old{In particular, records of communication events between individuals with high temporal resolutions have led to the study of dynamical properties of networks rather than static ones, in the emerging field of  temporal networks~\cite{Holme2012,Holme2015}.}
%\mycomment{First Referee, Minor \#1}
%\new{
In particular, records of communication events between individuals with high temporal resolutions have led to the study of dynamical properties of networks rather than static ones, in the emerging field of temporal networks (see Refs.~\cite{Holme2012,Holme2015,MasadaLambiotte2016} for comprehensive reviews).
In social temporal network studies, researchers have focused on the properties of the time series of interaction events associated to an individual (e.g., Refs.~\cite{Sanli2015,Borge-Holthoefer2015,Darst2016}).
%}
Such timelines of instantaneous events can be found in technology-mediated human communication, such as emails, short messaging service (SMS), or other message delivery services like Twitter, Facebook, google plus, and others.
Assuming that interaction events have a very short duration, as is often the case, this type of representation is also popular for mobile phone calls and physical proximity patterns measured by Bluetooth~\cite{Eagle2005} or RFID~\cite{Isella2011}.

%%% Literature review (Burstiness is an universal feature of human dynamics, based on IET distribution analysis)
Previous studies have  shown in a variety of domains that burstiness is a universal feature of time series of human interactions ~\cite{Eckmann2004,Barabasi:2005aa,Alexei2006PRE}.
The notion of burstiness refers to the bursting behavior of individuals, as they exhibit a high activity within short periods and occasionally exhibit long periods of silence.
Burstiness is often characterized by the fat tail of the distribution of the interevent times (IET) between successive events, sometimes fitted by power-law functions~\cite{Barabasi:2005aa}, and significantly deviating from the exponential distributions expected in the case of classical Poisson processes. These observations motivated the design of more elaborated models for human activity~\cite{Barabasi:2005aa,Malmgren2008}.

\begin{figure}[t]
\begin{center}
\includegraphics[]{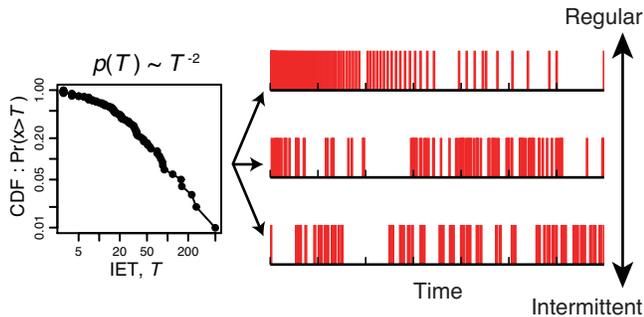}
\caption{(Color online) Three sequences of events with different temporal fluctuations (right) that are generated from the same set of the IETs (left).}
\label{fig:SchemeOfLV}
\end{center}
\end{figure}

%%% Problem statement (Analysis on temporal fluctuations in local time-domain is needed beyond the IET distribution analysis)
Although the presence of heavy-tailed IET distributions is a clear evidence of burstiness, the temporal dynamics of human communication is only partially described by a given IET distribution.
For example, as illustrated in Fig.~\ref{fig:SchemeOfLV}, a single set of heavy-tailed IETs yields event sequences with drastically different temporal fluctuations.
The sequence at the top panel looks like a regular IETs with slow frequency modulation.
In contrast, in the bottom sequence, the short and long IETs occur intermittently.
The middle one is in an intermediate situation.
Therefore, additional measures are required to distinguish these different behavior.
In fact, intermittent sequences associated to higher-order correlations between IETs have been reported in real-world examples~\cite{Goh2008,Karsai2012,Vestergaard2014}.
The characterization of such temporal fluctuations in social communication is an ongoing challenge to understand the nature of temporal dynamics of human communication.

%%%  Literature Review (Related discussions in neuroscience)
Similar questions are central in the field of computational neuroscience.
Characterization of the signal sequence of neuronal activity, called spike trains, is an important issue related to the problem of neural coding, which aims to understand how neurons communicate using spikes~\cite{dayan_abbott_2001tnc}.
A variety of methods have been developed for the analysis of spike trains \cite{gabbiani1998,kostal2007,Shinomoto:2003aa,Shinomoto2009}.
In the present work, we are especially interested in a measure originally designed to characterize temporal fluctuations in spike train data, called local variation (LV)~\cite{Shinomoto:2003aa,shinomoto2005,Shinomoto2009}.
The LV measure presents the advantage of being essentially orthogonal to the activity rate in the sense of information geometry~\cite{Miura:2006aa}, and has been shown to be a robust measure against the non-stationary modulation of the activity rate which were tested in multiple datasets in comparison with other statistical measures~\cite{shinomoto2005,Shinomoto2009}.
These properties make LV a promising candidate measure for the study of social communication data, as they are subject to external modulation of the activity rate, such as daily, weekly, and seasonally rhythms~\cite{Malmgren2008,Jo2012,KobaLambiotte2016}.
The LV measure was recently used to study the effect of popularity on  temporal fluctuations of events in Twitter~\cite{CeydaRenaud}.

% in- & out-relationship
An important difference between the previous studies and our work is to pay attention to the relationship between incoming and outgoing events involving social agents and its impacts on temporal fluctuations.
Similarly to neurons, receiving inputs and integrating them to send outputs, social agents are subject to incoming messages that may, or not, trigger reactions. In order to test this idea, we analyze the response-time distribution in empirical datasets and develop a generalized Hawkes process to model the observed dynamical properties.
The majority of previous studies on higher-order correlations between IETs~\cite{Goh2008,Karsai2012,Vestergaard2014} primarily focused on the event timings and dismissed the directions of messages.
However, investigation of the input-output relationship in human messaging processes may provide us important insight on how information flows in human communications. 

Toward this goal,  
in section~\ref{sec:characterization}, we introduce the LV measure and show that it provides a  characterization of temporal fluctuation of each individual. 
In section~\ref{sec:inout}, using the LV measure, we characterize the relationship between the incoming (receiving) and outgoing (sending) event sequences, and develop a point process model to identify the mechanism behind the observed correlations.
%\old{In section 4, we briefly examine the relationship between the  LV measure of users and   the conventional measures on the structural properties of social networks.}
Finally, in section~\ref{sec:discussion}, we summarize and discuss our findings.

%%%%%%%%%%%%%%%%
%%%%   Section 2. \section{Characterizing temporal fluctuations by statistical measures for a event sequence} 
%%%%%%%%%%%%%%%%
\section{Characterizing temporal fluctuations by statistical measures for event sequences}\label{sec:characterization}

\subsection{Datasets}\label{sec:data}
We analyze the social communication datasets of SMS, phone-calls, and emails. 
In the following, we refer to the three datasets as SMS, Phone, and Email.
The SMS and Phone datasets are a collection of timestamp of communication events made among a subset of anonymized users offered by a European cellphone service provider~\cite{Tabourier:2016ks}.
The SMS dataset contains 28,757,905 events among 983,424 unique users during one month and the Phone dataset contains 14,303,384 events among 1,131,049 unique users during the same period.
%\mycomment{Second Referee \#8}
%\new{
Although each event in the Phone dataset has a duration, we discard it and focus on the starting time in the following analysis. It should be noted that missed phone calls (with null duration) have been filtered out from the dataset.
%}
The Email dataset is Enron email network \cite{konect:klimt04} \footnote{http://konect.uni-koblenz.de/ (Date accessed: March 22nd, 2016).} that contains 1,148,072 emails sent between employees of Enron from 1999 to 2003. 
%\new{
This dataset was made public during the legal investigation concerning the Enron corporation. A user can send an email to oneself and we keep such self events for analysis.
%}
The time resolution of all the datasets is equal to one second.

\subsection{Statistical measure for event sequences}
%%%  Definitions
Let us consider a sequence of IETs denoted by $\left\{ T_1, T_2, \ldots, T_n \right\}$ between $n+1$ events, where $T_i$ is the length of $i$-th interval. Different statistical indicators can be used in order to characterize this time series. A popular choice is the coefficient of variation (CV), defined as
\begin{align}
\mathrm{CV} \equiv \frac{\sqrt{\frac{1}{n-1}\sum_{i=1}^{n} (T_i - \bar{T})^2 }} {\bar{T}},
\end{align}
where $\bar{T} \equiv \sum_i T_i / n$ is the average IET.
A large value of CV indicates the heterogeneity in the IETs. 
The CV is equivalent to the burstiness measure proposed in Ref.~\cite{Goh2008} up to algebraic variable transformation.

The local variation (LV) is instead defined by \cite{Shinomoto:2003aa}:
\begin{align}
\mathrm{LV} \equiv \frac{1}{n-1} \sum_{i=1}^{n-1} \frac{3 (T_i  - T_{i+1})^2}{(T_i  + T_{i+1})^2}, \label{eq:DefLV}
\end{align}
and mainly differs from CV by its comparison between  successive values of IETs.
The factor three of the numerator in the right hand side of Eq.~\eqref{eq:DefLV} is as to set the LV value for a Poisson process equal to unity. 
Theoretically, the LV value ranges from zero (i.e., regular) to three (i.e., intermittent).

%%% Theoretical result for Poisson process
Let us summarize some known properties of LV and its comparison with CV. 
The expected value of LV, denoted by ${\mathbb E}(\mathrm{LV})$, is equal to unity for a Poisson process with a constant activity rate~\cite{Shinomoto:2003aa}, and ${\mathbb E}({\rm CV})$ is also equal to unity in this case.
By the definition of LV in Eq.~\eqref{eq:DefLV}, if the fluctuations among successive IETs are smaller than those expected for a Poisson process, the LV value is smaller than unity.
In other words, the event sequence with ${\rm LV} < 1$ is more regular than a Poisson process.
By contrast, if the fluctuations are larger than those of a Poisson process and the event sequence is intermittent, the LV value is larger than unity.
When the event sequence is intermittent with a large LV value, we typically observe a negative correlation between two successive IETs; a short IET is likely to follow a long IET and vice versa.
In neuroscience literature~\cite{Shinomoto:2003aa,shinomoto2005,Shinomoto2009}, the intermittent sequence is referred as bursty,
because such an intermittent spike train is often recorded from bursting neurons, which is one of electrophysiological neuronal types~\cite{shepherd2004synaptic}.
It should be noted that burstiness is defined in a different way from the definition in the literature of network science~\cite{Barabasi:2005aa}.
% \old{Here, we use the term ``burst'' to refer to situations when LV has a value larger than unity, which does not imply a heavy-tail for the IET distribution, as often considered in in the literature. } \aoki{We do not use the term bursty. Use intermittent.}

The LV measure takes finite values even when the CV value diverges.
For example, let us suppose that the tail of the IET distribution exhibits a power-law function $\tau^{-\alpha}$.
If $\alpha \le 2$ holds, the CV value diverges to $+ \infty$ in the limit of a large number of events (and very large values in finite samples as shown in Fig.~\ref{fig:LVCV}).
Even in such a case, the LV value remains finite and can capture the correlations between successive IETs. 

\subsection{LV and CV values of empirical datasets}
\label{sec:LV_CV}

\begin{figure*}[t]
\begin{center}
\includegraphics[width=12cm]{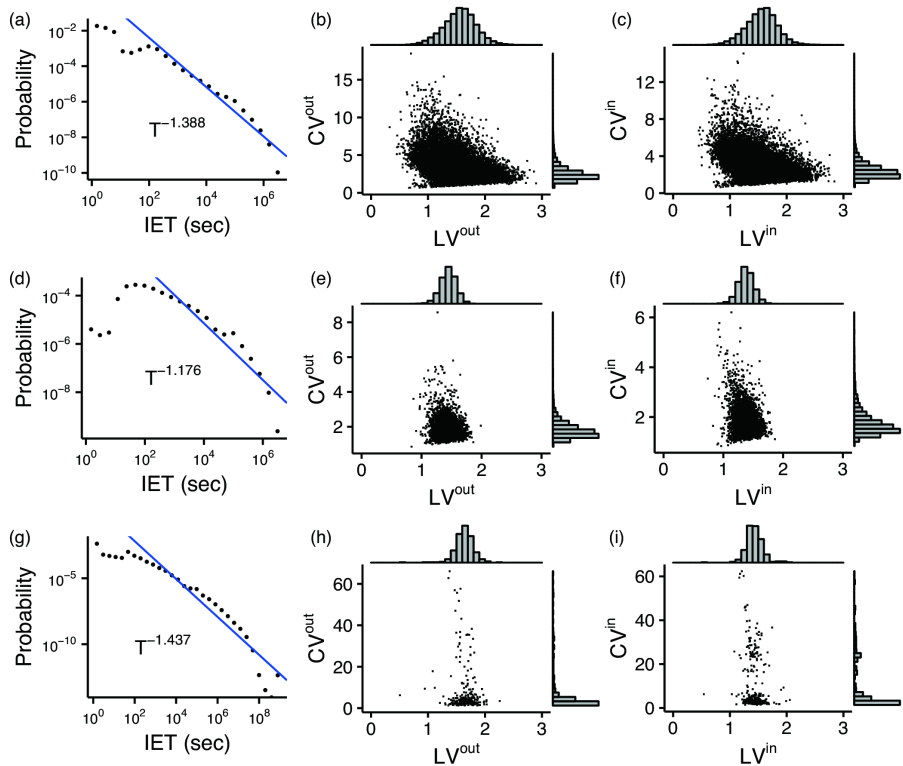}
\caption{(Color online) (a) Distribution of the IETs of outgoing events of the SMS dataset. Scatter plots of the LV and CV values for (b) outgoing events and (c) incoming events. The histograms of the measures are also shown.
The same set of the plots for (d)-(f) the Phone dataset and for (g)-(i) the Email dataset.}
\label{fig:LVCV}
\end{center}
\end{figure*}

We calculate the LV and CV values of users in the SMS, Phone, and Email datasets. 
The events in all of the three datasets have their directionality.
In other words, each user has two sequences of events: the incoming (i.e., receiving) events and the outgoing (i.e., sending) events of messages.
In this section, we separately evaluate these statistical measures for the incoming and outgoing events (i.e., $\CVin$, $\CVout$, $\LVin$, and $\LVout$) for each user.
In the following analysis, we restrict ourselves to consider the users having no less than $100$ IETs for both incoming and outgoing events so as to obtain stable results.
%\mycomment{Second Referee \#5}
%\new{
This restriction leaves the number of users to be analyzed equal to $50403$ ($5.1$ percent of the users in the original dataset) for SMS dataset, $4614$ ($0.41$\%) for Phone, and $290$ ($0.33$\%) for Email.
We adopted the threshold value (i.e., $100$ IETs) that was used in the previous studies~\cite{Shinomoto:2003aa,Shinomoto2009}.
In addition, as shown in Appendix~A, we numerically confirmed that the standard deviation of LV over synthetic event sequences with $100$ IETs is sufficiently small.
%}

% (Heterogenous IETs with large CV)
We plot the IET distribution of outgoing events for all the nodes in the SMS dataset (Fig.~\ref{fig:LVCV}(a)).
In accordance with the results reported in the previous studies on various communication data \cite{Barabasi:2005aa,Alexei2006PRE}, the IET distribution of outgoing events indicate a heavy-tailed behavior that roughly follows a power-law function. 
This heterogeneity in the outgoing IETs is also confirmed at the individual level; the $99.83\%$ of users have the $\CVout$ values larger than unity (Fig.~\ref{fig:LVCV}(b)).
% (LV distribution)
By contrast, the $\LVout$ values of users has a bell-shaped distribution, approximated by a normal distribution with the mean $1.56$ and standard deviation $0.28$. 
The range of the $\LVout$ values indicates the variety in message sending behavior; some individuals send messages in a regular manner and others send in a random or even intermittent manner.
In fact, the $97.1\%$ of users have the $\LVout$ values larger than unity, which implies that for those individuals, the IET fluctuations are larger than those of a Poisson process.
For the rest of users ($2.9\%$ of all), their temporal behavior are more regular than a Poisson process.
However, The $98.9\%$ of users with $\LVout < 1$ still have the $\CVout$ values larger than unity, indicating the heterogeneity of the set of IETs from these users (Fig.~\ref{fig:LVCV}(b)).
% (Correlation between LV and CV)
As we can see in Figs.~\ref{fig:LVCV}(b) and \ref{fig:LVCV}(c), there are no clear correlations between the CV and LV values of users for either incoming or outgoing events.
Therefore, the LV measure may capture characteristics of event sequences that cannot be explained by CV.
The same plots for the other two datasets are shown in Figs.~\ref{fig:LVCV}(d)-(i).
We can see similar behavior of LV and CV as in the SMS dataset.
The CV values in the Email dataset can be very large, possibly because of the long term of the observation (i.e., four years) and presence of very long IETs.

%%%  LV measures characterizes the personal temporal fluctuation stably (F-test)
\begin{table}[t]
\caption{Results of $F$-test statistics for the LV and CV values in the three datasets.}
\label{tbl:Ftest}
\begin{center}
\begin{tabular}{llrrr}
\hline
dataset & type  &  $0.1$-percentile & $F$-value (LV) & $F$-value (CV)\\
\hline
SMS  & incoming & 1.146 & 28.982 & 8.236\\
 & outgoing  & 1.137 & 32.740 & 9.011\\
Phone  & incoming  & 1.227 & 4.203 & 12.398\\
&  outgoing & 1.151 & 6.286 & 11.122 \\
Email & incoming & 1.178 & 3.377 & 1.160\\
& outgoing & 1.568 & 5.867 & 1.328\\
\hline
\end{tabular}
\end{center}
\end{table}%

In closing this section, we confirm the consistency of the LV and CV values of a user over time.
Because we want to use the LV and CV values as the steady characteristics of users, the variance of these values of a user across different periods must be smaller than the variance over the population.
We verify the consistency by using a statistical $F$-test~\cite{Wildt1978}, in which we compare the variances of the LV (CV) values across all nodes with the average of the variances of the LV (CV)  values of each node across $20$ subdivided sequences (see Appendix B for details).
As the results of the statistical tests summarized in Table~\ref{tbl:Ftest}, the $F$-values of the LV and CV values for both incoming and outgoing events are significantly above the $0.1$-percentile points for all of the three datasets, except for the CV values in the Email dataset.
These results indicate that the variance of the LV value of each user over time is significantly smaller than the variance over the population, and the usage of these measures as the users' characteristics is justified.

%% SECTION: Input-output relationship of individual human dynamics
\section{Relationship between incoming and outgoing event sequences} \label{sec:inout}

We first investigate the correlations between the statistics of  incoming and outgoing events of users
and study how individuals send messages in response to receiving messages.
Then, we propose a simple model for interpreting the observations in the datasets.

\subsection{Correlations between LV values of incoming and outgoing events}\label{sec:correlation_LV}
\begin{figure*}[tb]
\begin{center}
\includegraphics[width=12cm]{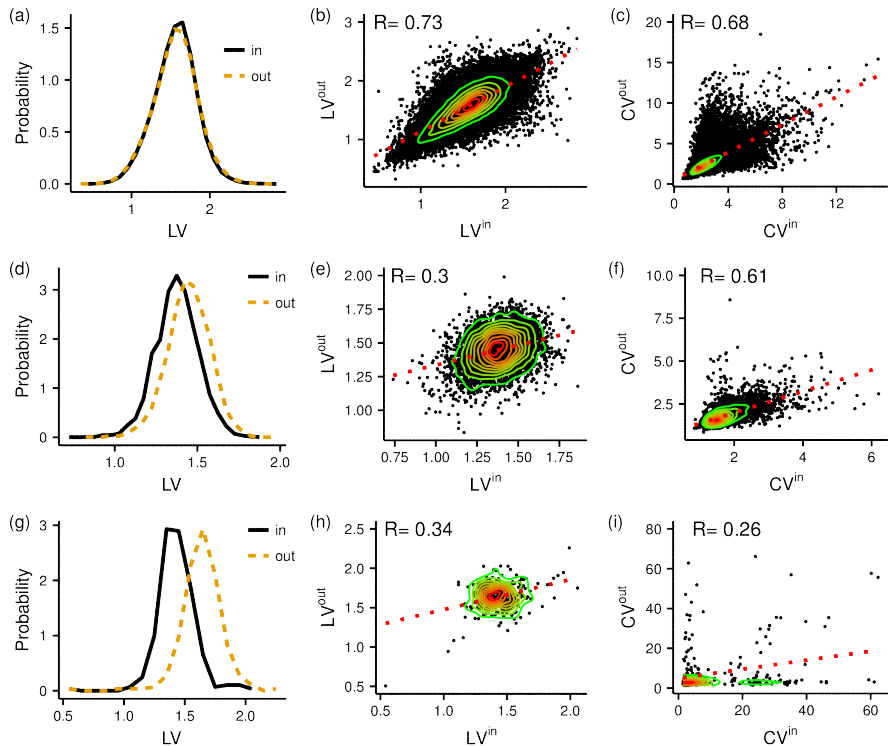}
\caption{(Color online) (a) Histogram of LVs, the correlation between (b) $\LVin$ and $\LVout$ and (c) $\CVin$ and $\CVout$ for the SMS dataset.
The same set of the plots for (d),(e),(f) the Phone dataset and (g),(h),(i) the Email dataset.
}
\label{fig:LVCorr}
\end{center}
\end{figure*}

%%% Problem Statement : How people response to receiving messages?
Figure~\ref{fig:LVCorr} shows the distributions of the LV values and the correlations of the LV (CV) values between incoming and outgoing events for the three datasets. 
For the SMS dataset, the $\LVin$ and $\LVout$ values are almost identically distributed (Fig.~\ref{fig:LVCorr}(a)). 
This identity of the two distributions is not trivial because these events are driven by different mechanisms as follows. 
On the one hand, receiving messages is a passive process for a user, because senders (i.e., other users) determine the timings.
If the actions of the senders are independent of each other as well as of the focal user's action, 
the correlations between the successive events are disappeared and 
the resultant event sequence of receiving messages can be modeled by a Poisson process with a time-dependent activity rate~\cite{Kass:2005go}. 
On the other hand, sending messages is an active process for a user, because the user determines the timing and it can differ from a Poisson process.
Therefore, this identity of the $\LVin$ and $\LVout$ distributions shown in Fig.~\ref{fig:LVCorr}(a) suggests that the event sequences of receiving messages from different senders are not independent of each other and may be correlated with the action of the receiver.

To examine the relationship of incoming and outgoing events at the individual level, we depict the scatter plots of the $\LVin$ and $\LVout$ values and the $\CVin$ and $\CVout$ values (Figs.~\ref{fig:LVCorr}(b) and \ref{fig:LVCorr}(c)).
The $\LVin$ and $\LVout$ values of a user exhibit a positive correlation, whereas the $\CVin$ and $\CVout$ values are less correlated.
%\mycomment{First Referee, Major \#1 and Second Referee \#4, \#6}
%\new{
The  95\% confidence interval of the correlation between $\LVin$ and $\LVout$ is equal to $[0.726,0.734]$ for SMS (Fig.~\ref{fig:LVCorr}(b)), while it is $[0.271, 0.323]$,  $[0.23,0.437]$ for Phone (Fig.~\ref{fig:LVCorr}(e)) and Email (Fig.~\ref{fig:LVCorr}(h)) datasets, respectively.
Thus, the correlation observed for SMS dataset significantly deviates from those for the other two datasets.
It should be noted that these correlation coefficients are robust against the change in the threshold value of the number of IETs that we use to filter the users (see Appendix~A).
%}

%\new{
Although the values of the Pearson correlation coefficients for the LV ($0.73$) and CV ($0.68$) plots are close, 
its 95\% confidence interval for the LV, $[0.726,0.734]$, deviates from that for the CV, $[0.679,0.689]$.
%}
The correlation between $\LVin$ and $\LVout$ values implies a possible interaction between the incoming and outgoing events, that is, a reaction behavior of users when replying to received messages. 

The Phone and Email datasets exhibit the $\LVin$ and $\LVout$ statistics different from those of the SMS dataset, while the results of the two datasets are similar (Figs.~\ref{fig:LVCorr}(d)--\ref{fig:LVCorr}(i)).
For the two datasets, the distributions of the $\LVin$ and $\LVout$ values (Figs.~\ref{fig:LVCorr}(d) and \ref{fig:LVCorr}(g)) are less similar than those for the SMS dataset (Figs.~\ref{fig:LVCorr}(a)).
In addition, the correlations between the $\LVin$ and $\LVout$ values (Figs.~\ref{fig:LVCorr}(e) and \ref{fig:LVCorr}(h)) are much weaker than that for the SMS dataset (Figs.~\ref{fig:LVCorr}(b)).
%\mycomment{Second Referee \#7}
%\new{
For the Phone dataset, the correlation for LV is also much weaker than that for CV.
%}

These differences in the LV statistics between the three datasets may be owing to the different communication manner for these communication tools. 
For example, users of SMS may quickly respond to a received message.
For phone-call communication, such quick response (i.e., back call) may not be necessary because one can have conversations within a single call.
In a similar way, for email communication, many messages are left in mailbox and later the replies to them are sent.
To examine the response behavior between the datasets, we will employ the response-time distribution analysis in the next section.

%%%%%%%%%%%%%%%%%%%%%%%%%%%%%%%%%%%%
%%%  SUBSECTION: Introduction of Reponse-time distribution
%%%%%%%%%%%%%%%%%%%%%%%%%%%%%%%%%%%%%
\subsection{Response behavior to incoming messages}\label{sec:eta}

\begin{figure}[t]
\begin{center}
\includegraphics[]{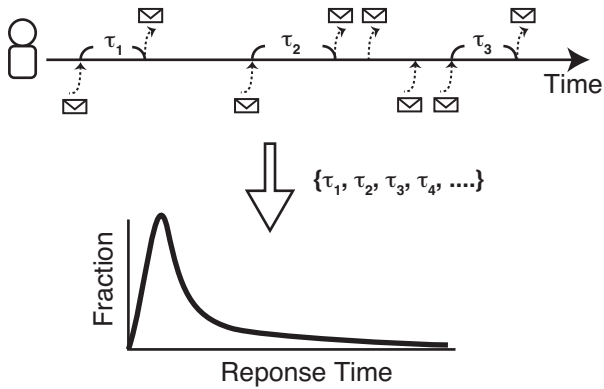}
\caption{(Color online) Schematic of computation of the response-time distribution.}
\label{fig:ETAIllust}
\end{center}
\end{figure}

\begin{figure*}[t]
\begin{center}
\includegraphics[width=14cm]{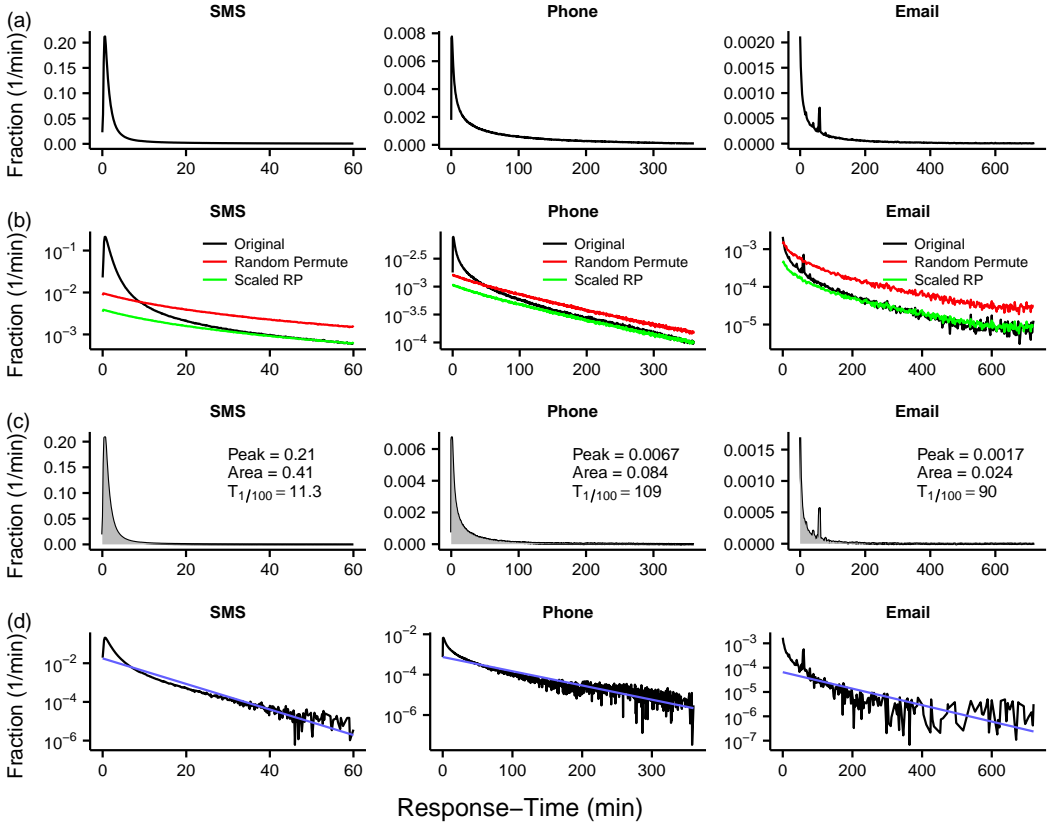}
\caption{
(Color online) Response-time distributions for the SMS, the Phone, and the Email datasets.
(a) Normal plots.
(b) Comparisons with the response-time distribution obtained from the randomized datasets. 
(c) The net response functions to an incoming event after subtracting the null-model baseline.
(d) Semi-log plots fitted by an exponential function.
}
\label{fig:ETAResult}
\end{center}
\end{figure*}

To clarify the differences between the three datasets in terms of the response patterns, 
the response-time distributions are computed for the SMS, the Phone, and the Email datasets. 
The procedure for calculating a response-time distribution is schematically shown in Fig~\ref{fig:ETAIllust}.
We define the response time to an incoming event as the time interval until the first outgoing event of the user (top of Fig.~\ref{fig:ETAIllust}). 
We count the response time if and only if there is no other incoming event between the focal pair of the incoming and outgoing events, so as to avoid the duplication of the response time.
The response-time distribution is then computed by constructing a histogram of the response times for all the users (bottom of Fig.~\ref{fig:ETAIllust}). 
The bin size of the histogram is set to $10$, $20$, and $120$ seconds for the SMS, the Phone, and the Email datasets, respectively.
The bin size for the Email dataset is larger than those for the other datasets so as to remove the peaks of activities at every minute which may be artefact due to batch mail delivery system. 
%   1) Simple observations
As shown in Fig.~\ref{fig:ETAResult}(a), the response-time distributions of all the datasets have a sharp peak just after an incoming event and decreases with time. 
%\mycomment{First Referee, Minor \#2}
The distribution for all the datasets have sharp peaks at $\tau \sim 1$ minute.
It should be noted that a similar peak is also reported in Ref. \cite{Saramaki2015PRE} although a different definition of the response time is adopted
Only the Email dataset has another smaller peak at $\tau \sim 1$ hour.

%%% Result of the ETA analysis

%%%  Comparison with a randomized data 
Although the response-time distribution (Fig.~\ref{fig:ETAResult}(a)) provides full information of the response behavior in the datasets, 
we want to know which part of the response-time distribution cannot be explained by a baseline activity patterns, or a null model, of individuals.  
To generate such a null-model event sequences, we shuffle the time stamps of all the events in a dataset (this shuffling is called randomly permuted times~\cite{PhysRevE.71.046119,Holme2012}).
This randomization destroys any temporal correlations between the event times, while retaining the total number of incoming and outgoing events of each user and the total number of events for each user pair. 
The randomization also conserves the total number of events occurring at each time and consequently daily and weekly activity patterns.

We calculate the response-time distribution for the randomized dataset and compare them with the original ones (Fig.~\ref{fig:ETAResult}(b)).
The response-time distributions for the randomized datasets do not have the apparent peaks, in contrast to the original response-time distributions, while they exhibit decay for large $\tau$ similar to the original distributions.
If we vertically rescaled the null-model response-time distributions, they mostly overlaps with the original distributions for large $\tau$.
This result implies that the decay of the distribution is not related to the response behavior of users, because the randomization destroys any temporal correlations between the incoming and outgoing events.

Subtracting the rescaled null-model curves from the original response-time distributions reveals the response behavior of users that deviate from the null model (Fig.~\ref{fig:ETAResult}(c)).
We refer to the resultant curves as the net response functions to an incoming event.
The impacts of an incoming event are evaluated by the largest peak  of the response function and the area under it.
The peak and area are equal to $0.23$ and $0.41$ for the SMS dataset, $0.0067$ and $0.084$ for the Phone dataset, and $0.0017$ and $0.024$ for the Email dataset.
These results suggest that the impacts of an incoming event is much stronger in the SMS dataset than those in the other two datasets. 
We also quantify the decay of the net response function by the time required for the net response function to decay to the $1/100$ value of the peak after the peak time, denoted by $T_{1/100}$.
We decide to measure the decay in this way because the net response functions cannot be well fitted by an exponential function (Fig.~\ref{fig:ETAResult}(d)).
The $T_{1/100}$ values are equal to $11.3$, $109$, and $90$ minutes for the SMS, the Phone, and the Email datasets.
These results indicate that the response of users in the SMS dataset is much faster than those in the Phone and the Email datasets.

%%%%%%%%%%%%%%%%%%%%%%%%%%%%%%%%%%%%%%%%%%%%%%%%
%%
%% Subsection: Interpret the differential property in the correlation between LV(in) and LV(out) by using a modified Hawkes process model
%%
%%%%%%%%%%%%%%%%%%%%%%%%%%%%%%%%%%%%%%%%%%%%%%%%%%%%

\subsection{A generalized Hawkes model incorporating response behavior}\label{sec:Hawkes}

%%% The aim of this section
Combining the results described in the previous sections, we are interested in the relationship between 
the correlations between the $\LVin$ and $\LVout$ values (Fig.~\ref{fig:LVCorr}) and the response behavior of users (Fig.~\ref{fig:ETAResult}). 
The hypothesis is that the fast and intense response such as observed in the SMS dataset induces chain reactions of messaging events between individuals in a short period and thus a positive correlation between $\LVin$ and $\LVout$ emerges. 
By contrast, the slow and moderate response observed in the Phone and the Email datasets do not cause a strong correlation between the LV values.  
To validate this hypothesis, we introduce a point process model of communication activities and examine the dependence of the correlation between the $\LVin$ and $\LVout$ values on the response behavior. 

%%% Model description
Our model is based on the Hawkes process~\cite{Hawkes}, which was first proposed to model seismic patterns~\cite{ogata1988} and was recently applied to human communication behavior~\cite{masuda2013,PhysRevE.89.042817,SEISMIC,KobaLambiotte2016},
In the Hawkes model, the activity rate of an individual, denoted by $\lambda(t)$, is determined by
\begin{align}
\lambda(t) = u_0 + a_{\self} \sum_{k: t_k^{{\rm out}} < t}  \frac{1}{\tau_{\self}} \exp \left[ - \left( \frac{t - t_k^{\mathrm{out}}}{\tau_{\self}} \right) \right], \label{eq:Hawkes}
\end{align}
where $u_0$ is the baseline activity rate, $t_k^{\rm out}$ is time of the $k$-th outgoing event and $a_{\self}$ and $\tau_{\self}$ are the amplitude and the time constant of self-modulation (i.e., effects of outgoing events). 
We incorporate the effect of response behavior to incoming events from others (i.e., receiving messages) by adding the corresponding term to Eq.~\eqref{eq:Hawkes} as:
\begin{align}
\lambda(t) = u_0 
 + a_{\self} \sum_{k: t_k^{\mathrm{out}} < t}  \frac{1}{\tau_{\self}} \exp \left[ - \left( \frac{t - t_k^{\mathrm{out}}}{\tau_{\self}} \right) \right] \notag \\
 + a_{\ex} \sum_{k: t_k^{\mathrm{in}} < t} \frac{1}{\tau_{\ex}}\exp \left[ - \left( \frac{t - t_k^{\mathrm{in}}}{\tau_{\ex}} \right) \right], \label{eq:model}
\end{align}
where $t_k^{\mathrm{in}}$ is the time of the $k$-th incoming event and $a_{\ex}$ and $\tau_{\ex}$ are the amplitude and the time constant of the response behavior.

%%%%%%%%%%%%%%%%%%%%%%%%%
%%% Simulation setup
%%%%%%%%%%%%%%%%%%%%
The incoming event sequence represents an exogenous effect of the model and we have to set its generation process for numerical simulations.
As shown in Figs.~\ref{fig:LVCV}(c,f,i), the $\LVin$ values follow a bell-shaped distribution.
To mimic this situation, we first randomly draw a $\LVin$ value from the normal distribution with mean $\mu = 1$ and the standard deviation $\sigma = 0.25$.

Then, we generate an event sequence with the given $\LVin$ value by using a gamma process.
A gamma process is a renewal process whose IETs, denoted by $T$, obey a gamma distribution:
\begin{align}
p(T; \kappa, \theta) = T^{\kappa -1} \frac{e^{-T/\theta}}{\theta^\kappa \Gamma(\kappa)},
\end{align}
where $\kappa$ and $\theta$ are parameters and $\Gamma(x)$ is the gamma function.
The mean and variance of IETs and the LV value of the gamma process are given by~\cite{Shinomoto:2003aa}:
\begin{align}
 {\mathbb E}[T] = \kappa \theta, \quad {\rm Var}[T] = \kappa \theta^2, \quad {\rm LV} = \frac{3}{2\kappa + 1}.
\end{align}
Therefore, we set $\kappa = (3/\LVin -1)/2$ to achieve the given $\LVin$ value and then $\theta = 1/\kappa$ to retain $ {\mathbb E}[T] = \kappa \theta = 1$.
For a drawn $\LVin$ value, we run a numerical simulation of the model of Eq.~\eqref{eq:model} until we obtain $100$ IETs~\footnote{In the empirical data analysis, the sequence that has at least $100$ IETs are selected to evaluate LV and CV measures (see section~\ref{sec:LV_CV}).}.
Then, we compute the $\LVout$ value of the generated event sequence.

%%%%%%%%%%%%%%%%%%%%%%%%%%%
%%
%%% Results
%%
%%%%%%%%%%%%%%%%%%%%%%%%%%%

\begin{figure*}[t]
\begin{center}
\includegraphics[]{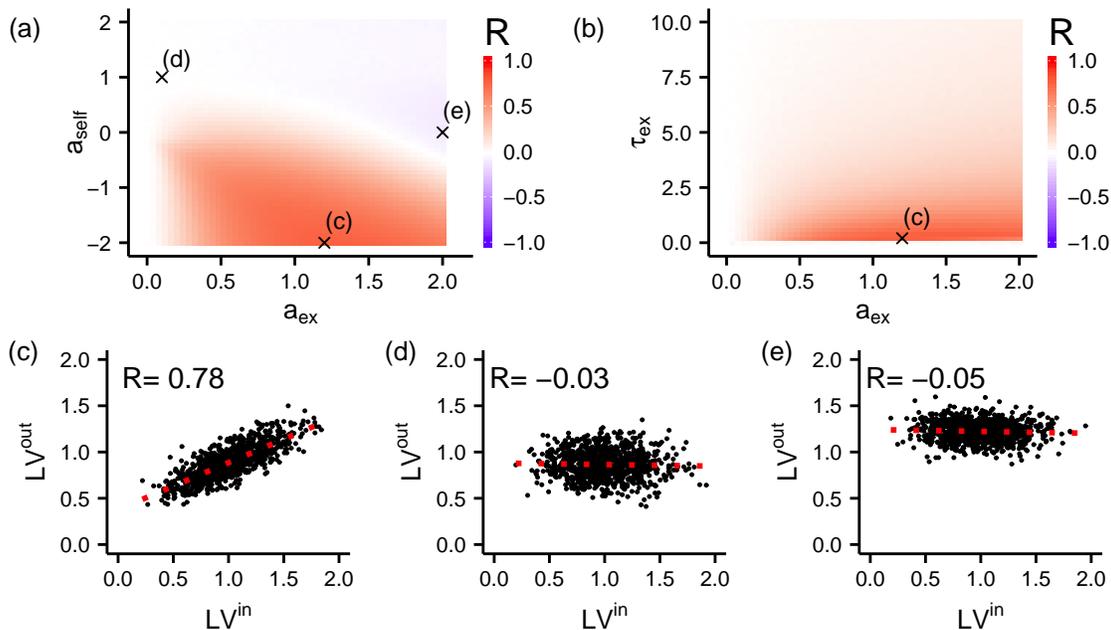}
\caption{
(Color online) (a),(b) Phase diagrams of the correlation coefficient $R$ between the $\LVin$ and $\LVout$ values obtained by the numerical simulations of the model (Eq.~\eqref{eq:model}).
The parameters that are not indicate by the axis are set to $(u_0, a_{\self}, \tau_{\ex}, \tau_{\self}) = (0.2, -2, 0.2, 0.5)$.
Scatter plots of the $\LVin$ and $\LVout$ values obtained with (c) $(u_0, a_{\ex}, a_{\self}, \tau_{\ex}, \tau_{\self}) = (0.2, 1.2, -2, 0.2, 0.5)$, (d) $(0.2, 0.1, 1, 0.2, 0.5)$, and (e) $(0.2, 2, 0, 0.2, 0.5)$.
These parameter settings are indicated by cross points in the diagrams (a) and (b).
}
\label{fig:HawkesModel}
\end{center}
\end{figure*}

We summarize the numerical results of the model into the phase diagrams of the Pearson correlation coefficient $R$ between $\LVin$ and $\LVout$ in Figs.~\ref{fig:HawkesModel}(a) and \ref{fig:HawkesModel}(b).
We draw randomly $100000$ $\LVin$ values and carry out the simulations for a fixed set of parameters $(u_0, a_{\self}, \tau_{\ex}, \tau_{\self})$. 
%\mycomment{First Referee, Major \#1 and Second Referee \#4}
These diagrams indicate that the positive correlation between the $\LVin$ and $\LVout$ values, which is similar to that observed in the SMS dataset (Fig.~\ref{fig:LVCorr}(b)), emerges if $a_{\ex}$ is positively large and $a_{\self}$ is negatively large and $\tau_{\ex}$ is small.
A typical scatter plot with 1000 points in this condition is shown in Fig.~\ref{fig:HawkesModel}(c).
This parameter setting can be interpreted as follows.
The combination of a large $a_{\ex}$ and a small $\tau_{\ex}$ implies that the impact of an incoming event is strong and its response is very quick, as the response behavior observed in the SMS dataset (Fig.~\ref{fig:ETAResult}(c)).
Another key factor is a negative $a_{\self}$, which represents a refractory effect that the activity rate decreases after sending a message.
This refractory effect might be interpreted as  a user who just sent a message is satisfied and stop further sending or that 
some interval is required to write the next message.
In addition, $\tau_{\ex} < \tau_{\self}$ holds and the impact of incoming events decays faster than that of outgoing events.
Without these conditions, the $\LVin$-$\LVout$ correlation is not observed as shown in Figs.~\ref{fig:HawkesModel}(d) and \ref{fig:HawkesModel}(e).
In one case, 
when the model user has a weak response behavior  (e.g., $a_{\ex} = 0.1$) and a strong self-excitation effect (e.g., $a_{\self} = 1$),
there is no correlation between the LV values (Fig.~\ref{fig:HawkesModel}(d)).
In another case, when the model user has a strong response behavior (e.g, $a_{\ex}$ = 2) and does not have self-modulation effect (i.e., $a_{\self} = 0$),  the correlation is also almost equal to zero (Fig.~\ref{fig:HawkesModel}(e)).

On the basis of these results, the working hypothesis should be modified as follows.
The intense and quick response to incoming events (i.e., a large $a_{\ex}$ and a small $\tau_{\ex}$) and the refractory effect (i.e., a negative $a_{\self}$) are the fundamentals of the positive correlation between the $\LVin$ and $\LVout$ values.

\section{Discussion}\label{sec:discussion}
%% Summary of this study
In this paper, we have studied the temporal characteristics of human communication behavior using by the spike train analysis techniques.
%\mycomment{First Referee, Major \#1 and Second Referee \#4}
First, we have introduced LV measure to evaluate how the incoming and outgoing event intervals are temporally fluctuating, and found the positive correlation between $\LVin$ and $\LVout$ for the SMS dataset, while little correlation for the Phone and Email datasets.
Second, we have analyzed the response time of users to quantify how individuals send messages in response to receiving messages.
The comparison of the net response-time function for the original and the randomized datasets have unveiled a strong and quick response in the SMS dataset, contrary to the weak and slow response in the Phone and Email datasets.
To understand the mechanism behind these observations, we have developed a point process model based on the Hawkes model.
From this model study, we identified that the positive LV correlation can be reproduced by two key factors: a strong and quick response to incoming events and a refractory period after outgoing events.
%\old{In addition, we have studied the relationship between temporal and structural properties, and found that LV measures are almost uncorrelated with the conventional structural measures of the node, except weak correlations with the node strength.}

%%%%%%%%%%%%%
%%%
%%%  Comparison with Karsai et al. Sci. Rep 2012 (cited here) should be carefully discussed in Discussion section. by Taro
%%%
%%%%%%%%%%%%%
It is worth noting the difference between the present study and previous works.
We used the LV measure to capture the correlations between successive IETs in this study.
This is a way to quantify higher-order correlations in the event sequences beyond the statistics of single IETs.
Some previous work also considered such higher-order correlations and different ways of their characterizations have been discussed~\cite{Goh2008,Karsai2012,Vestergaard2014}.
An important difference between the present study and others is the attention to the input-output relationship in social communications.
For example, in Ref.~\cite{Goh2008}, the correlation of IETs is measured by the Pearson correlation coefficient of two successive IETs.
In Ref.~\cite{Karsai2012}, the correlation and memory effect in event sequences have been discussed by counting the number of events occurring within a time window.
However, the directions of contacts were omitted in both of the two studies.
Combination of the event counting statistics and the input-output relationship analysis, for instance, would provide new insight in understanding the higher-order correlations hidden in human social communications.

%%%%%%%%%%%%%
%%%
%%%  Discussion about the response time function
%%%
%%%%%%%%%%%%%
% the following line was commented out to kill a TeX compile error.
%\mycomment{First Referee, Major \# 2 and Second Referee \# 3}
%\new{
In section~\ref{sec:eta}, we defined the response time to an incoming event as the time interval until the first outgoing event made by the user after the incoming event.
This definition gives the lowest estimation of the true response time.
Context analysis of message contents may give a closer estimation, however, achieving a perfect matching between messages is still challenging.
Furthermore, the message contents of social communications are often not accessible for privacy protection.
Therefore, we decided to use the simplest estimation of the response time based on available data.
In a preliminary analysis, we checked the receivers of the outgoing events made by a user and found that most of receivers are the sender of the last incoming event to the focal user.
In other words, most of the outgoing events were the return calls or messages to the users who made the latest incoming event.
This observation supports our assumption behind the definition of the response time, although it does not completely exclude possible underestimates.
%}

%\new{
The timing between the arrival of a message and its notification to the user is another channel-dependent  factor which may have impacts on the response time but was not taken into account in this study.
We may frequently receive and check the notifications of incoming SMS messages and Phone calls on mobile phones. By contrast, we may less frequently check incoming emails on computers than we do on mobile phones (this may result in a small peak in the response function at around one hour for Email dataset, as shown in Fig.~\ref{fig:ETAResult}).
This direction of analysis requires the observation of the time when users recognize incoming messages, which is not included in the three datasets we considered.
Other datasets with such observation, if any, would lead to further understanding of the social response behavior in future work.
%}

%%%%%%%%%%%%%%%%%%%%%%%%%%%%
%%%
%%% Unsolved Problem 1 : The in/out LV correlation is caused by the type of communication method?
%%% Unsolved Problem 2 : Future studies on the input-output relationship will lead to the he understanding of the human dynamics.
%%%
%%%%%%%%%%%%%%%%%%%%%%%%%%%%%%%%%
The present study leaves several open research questions.
The first question is to clarify the relationship between the type of communication channel and the properties of the event sequences.
Although we observed different response behavior of users in several datasets, it is not clear whether the response behavior is common for the same type of communication (e.g., SMS) or is unique for the dataset used in this study. 
A comparison analysis of different instances of datasets for the same communication type would provide an answer to this question.
The second question, more general one, is to develop better stochastic models to describe the input-output relationship in social communications.
Our proposed model has a limitation that it does not distinguish source nodes of incoming events nor target nodes of outgoing events.
In reality, there should be heterogeneity in communication patterns between different user pairs.
In addition, our model does not explicitly consider the structure of social networks. 
Aside from the limitation of models, another important issue is to develop a procedure for evaluating the models. 
In computational neuroscience, proper benchmarks of the goodness of dynamical models has encouraged researchers to develop better models that can reproduce the input-output relationship observed in actual neuronal data~\cite{jolivet2008,MATMODEL}.
Similarly, introducing adequate benchmarks for social communication datasets would help us for further understanding of human dynamics on the basis of quantitative models.  

After the acceptance of this manuscript for publication in a journal, we noticed that the analysis of response times similar to our approach had been done in Ref. \cite{Saramaki2015PRE}.

%%%%%%%%%%%%%%%%%%%%%%%%%%%
% ACKNOWLEDGEMENTS
%%%%%%%%%%%%%%%%%%%%%%%%%%%
\section*{Acknowledgments}
The authors thank Shigeru Shinomoto, Rapha\u{e}l Li\'{e}geois, and Paul Expert for fruitful discussions.
The SMS and Phone dataset was provided by Lionel Tabourier.
The Email dataset was download from The Koblenz Network Collection.
This work was partly supported by Bilateral Joint Research Program between JSPS and F.R.S.-FNRS.
T.A.  acknowledges financial support from JSPS KAKENHI Grant Number 24740266 and 26520206. 
R.K. acknowledges financial support from JSPS KAKENHI Grant Number 25870915.
R.L. acknowledges financial support from ARC and the Belgian Network DYSCO (Dynamical Systems, Control, and Optimismtion), funded by the Interuniversity Attraction Poles Programme.

%%%%%%%%%%%%%%
%% Appendix
%%%%%%%%%%%%%
\appendix

%\new{
\section{Relationship between Local Variation and the number of IETs}
\begin{figure}[t]
  \begin{center}
	\includegraphics[width=7cm]{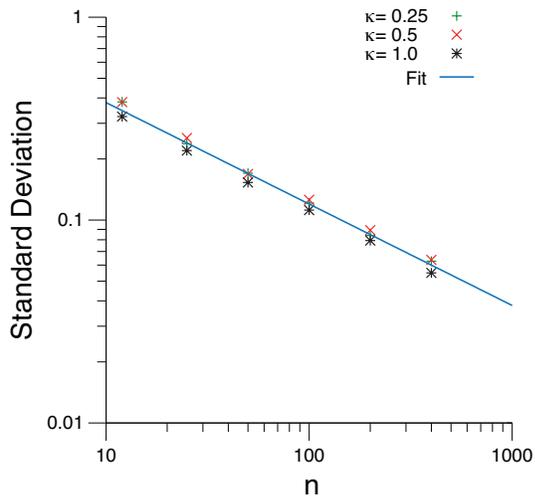}	   
	\vspace{0.5cm}	
	\caption{(Color Online) Effect of the number of IETs on the standard deviation of the LV values. For each point, the standard deviation was calculated over 100 instances of event sequences generated from the Gamma process with a given number of IETs. The results were well fitted with a power-law function $1.2 n^{-0.5}$.}
         \label{fig:Uncert_Lv}
  \end{center}
\end{figure}

The LV values fluctuate over different instances of event sequence with the same number of IETs. 
We numerically investigated the dependency of the standard deviation of the LV values on the number of IETs for the synthetic event sequences generated from the Gamma process (see section~\ref{sec:Hawkes} for its definition).
Figure \ref{fig:Uncert_Lv} shows the standard deviation of the LV values over different instances as a function of the number of IETs. 
Regardless of the parameter values of the Gamma process, the standard deviation algebraically decreases with the number of IETs and reaches to a sufficiently small value $(\sim 0.1)$ for $100$ IETs.
%}    

\begin{figure*}
  \begin{center}
	\includegraphics[]{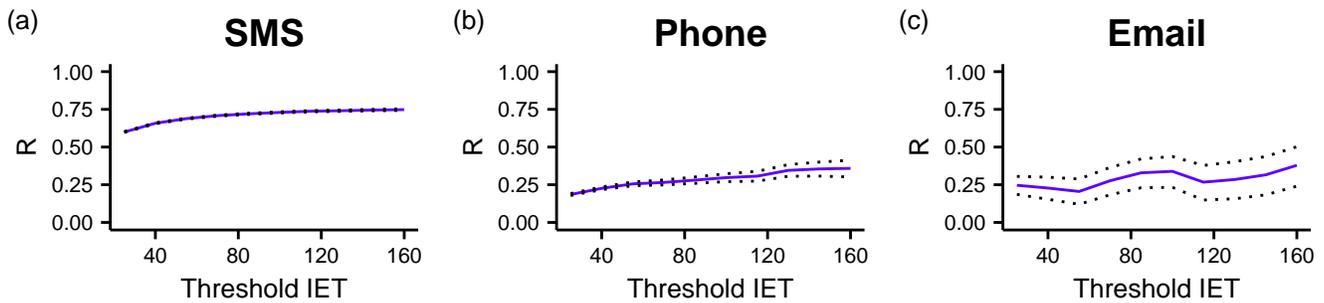}	   
	%\vspace{0.5cm}	
	\caption{(Color Online) Correlation coefficient $R$ between $\LVin$ and $\LVout$ of the users who have both incoming and outgoing events no less than the threshold value of the number of IETs. The dotted lines show the $95\%$ confidence intervals of the $R$ values.}
         \label{fig:Threshold}
  \end{center}
\end{figure*}

%\new{
According to the obsevation described in the previous paragraph, we set the threshold value of the number of IETs equal to $100$ to filter users when we calculated the LV values in section~\ref{sec:LV_CV} and the following.
We confirmed that the correlation coefficients between $\LVout$ and $\LVin$ reported in Fig.~\ref{fig:LVCorr} are robust against the change in the threshold value (see Fig.~\ref{fig:Threshold}).
%}

\section{Statistical $F$-test (analysis of variance)}
To evaluate how the measures (LV and CV) worked in discrimination of individual temporal patterns,
we conducted the $F$-test ~\cite{Wildt1978}, which compares the variance of the measure means across $N$ users to the mean of the measure variances across $n$ (=20) fractional sequences of individuals.
The null hypothesis of this test is
$$
	\mu_1 = \mu_2 = \cdots = \mu_N,
$$
where $\mu_i$ is the mean of measure (LV and CV) of user $i$ across $n$ fractional sequences.

For a given set of LV  values $\{\LV_{ij}\}$, each of which is computed for the $j$-th fractional IET sequence ($j$ = $1, 2, \cdots, n$) of  user $i$ ($i$ = $1, 2, \cdots, N$), the $F$-value is given by
\begin{align}
 F = \frac{ n \frac{1}{N-1}\sum_{i=1}^{N} \left(\overline{\LV}_i - \langle \overline{\LV} \rangle \right)^2} 
 {\frac{1}{N} \sum_{i=1}^N  \frac{1}{n-1} \sum_{j=1}^{n} (\LV_{ij}  - \overline{\LV}_i)^2},
 \end{align}
where $M = Nn$ is the size of the set  of  values $\{\LV_{ij}\}$ and $\overline{\LV}_i  = \sum_{j=1}^{n} \LV_{ij} /n$, 
$\langle \overline{\LV} \rangle = \sum_{i=1}^{N} \overline{\LV}_{i} /N$.
This $F$-value follows the $F$-distribution with $N - 1$, $M - N$ degrees of freedom under the null hypothesis.
The same test has been performed for CV.
The subset of users who have no leas than 1000 IETs are analyzed in this $F$-test, because each fractional sequence have no less than 50  IETs to evaluate LV and CV.

%%%%%%%%%%%%%%
%% bibliography
%%%%%%%%%%%%%

%\bibliography{main,neural}

\begin{thebibliography}{47}%
\makeatletter
\providecommand \@ifxundefined [1]{%
 \@ifx{#1\undefined}
}%
\providecommand \@ifnum [1]{%
 \ifnum #1\expandafter \@firstoftwo
 \else \expandafter \@secondoftwo
 \fi
}%
\providecommand \@ifx [1]{%
 \ifx #1\expandafter \@firstoftwo
 \else \expandafter \@secondoftwo
 \fi
}%
\providecommand \natexlab [1]{#1}%
\providecommand \enquote  [1]{``#1''}%
\providecommand \bibnamefont  [1]{#1}%
\providecommand \bibfnamefont [1]{#1}%
\providecommand \citenamefont [1]{#1}%
\providecommand \href@noop [0]{\@secondoftwo}%
\providecommand \href [0]{\begingroup \@sanitize@url \@href}%
\providecommand \@href[1]{\@@startlink{#1}\@@href}%
\providecommand \@@href[1]{\endgroup#1\@@endlink}%
\providecommand \@sanitize@url [0]{\catcode `\\12\catcode `\$12\catcode
  `\&12\catcode `\#12\catcode `\^12\catcode `\_12\catcode `\%12\relax}%
\providecommand \@@startlink[1]{}%
\providecommand \@@endlink[0]{}%
\providecommand \url  [0]{\begingroup\@sanitize@url \@url }%
\providecommand \@url [1]{\endgroup\@href {#1}{\urlprefix }}%
\providecommand \urlprefix  [0]{URL }%
\providecommand \Eprint [0]{\href }%
\providecommand \doibase [0]{http://dx.doi.org/}%
\providecommand \selectlanguage [0]{\@gobble}%
\providecommand \bibinfo  [0]{\@secondoftwo}%
\providecommand \bibfield  [0]{\@secondoftwo}%
\providecommand \translation [1]{[#1]}%
\providecommand \BibitemOpen [0]{}%
\providecommand \bibitemStop [0]{}%
\providecommand \bibitemNoStop [0]{.\EOS\space}%
\providecommand \EOS [0]{\spacefactor3000\relax}%
\providecommand \BibitemShut  [1]{\csname bibitem#1\endcsname}%
\let\auto@bib@innerbib\@empty
%</preamble>
\bibitem [{\citenamefont {Wasserman}\ and\ \citenamefont
  {Faust}(1994)}]{wasserman}%
  \BibitemOpen
  \bibfield  {author} {\bibinfo {author} {\bibfnamefont {S.}~\bibnamefont
  {Wasserman}}\ and\ \bibinfo {author} {\bibfnamefont {K.}~\bibnamefont
  {Faust}},\ }\href@noop {} {\emph {\bibinfo {title} {Social Network Analysis:
  Methods and Applications}}}\ (\bibinfo  {publisher} {Cambridge University
  Press},\ \bibinfo {year} {1994})\BibitemShut {NoStop}%
\bibitem [{\citenamefont {Lazer}\ \emph {et~al.}(2009)\citenamefont {Lazer},
  \citenamefont {Pentland}, \citenamefont {Adamic}, \citenamefont {{Sinan
  Aral}}, \citenamefont {Barab{\'{a}}si}, \citenamefont {Brewer}, \citenamefont
  {Christakis}, \citenamefont {Contractor}, \citenamefont {Fowler},
  \citenamefont {Gutmann}, \citenamefont {Jebara}, \citenamefont {King},
  \citenamefont {Macy}, \citenamefont {Roy},\ and\ \citenamefont
  {Alstyne}}]{Lazer2009}%
  \BibitemOpen
  \bibfield  {author} {\bibinfo {author} {\bibfnamefont {D.}~\bibnamefont
  {Lazer}}, \bibinfo {author} {\bibfnamefont {A.}~\bibnamefont {Pentland}},
  \bibinfo {author} {\bibfnamefont {L.}~\bibnamefont {Adamic}}, \bibinfo
  {author} {\bibnamefont {{Sinan Aral}}}, \bibinfo {author} {\bibfnamefont
  {A.-L.}\ \bibnamefont {Barab{\'{a}}si}}, \bibinfo {author} {\bibfnamefont
  {D.}~\bibnamefont {Brewer}}, \bibinfo {author} {\bibfnamefont
  {N.}~\bibnamefont {Christakis}}, \bibinfo {author} {\bibfnamefont
  {N.}~\bibnamefont {Contractor}}, \bibinfo {author} {\bibfnamefont
  {J.}~\bibnamefont {Fowler}}, \bibinfo {author} {\bibfnamefont
  {M.}~\bibnamefont {Gutmann}}, \bibinfo {author} {\bibfnamefont
  {T.}~\bibnamefont {Jebara}}, \bibinfo {author} {\bibfnamefont
  {G.}~\bibnamefont {King}}, \bibinfo {author} {\bibfnamefont {M.}~\bibnamefont
  {Macy}}, \bibinfo {author} {\bibfnamefont {D.}~\bibnamefont {Roy}}, \ and\
  \bibinfo {author} {\bibfnamefont {M.~V.}\ \bibnamefont {Alstyne}},\
  }\href@noop {} {\bibfield  {journal} {\bibinfo  {journal} {Science}\ }\textbf
  {\bibinfo {volume} {323}},\ \bibinfo {pages} {721} (\bibinfo {year}
  {2009})}\BibitemShut {NoStop}%
\bibitem [{\citenamefont {Albert}\ and\ \citenamefont
  {Barabasi}(2002)}]{Albert:2002p869}%
  \BibitemOpen
  \bibfield  {author} {\bibinfo {author} {\bibfnamefont {R.}~\bibnamefont
  {Albert}}\ and\ \bibinfo {author} {\bibfnamefont {A.}~\bibnamefont
  {Barabasi}},\ }\href {\doibase 10.1103/RevModPhys.74.47} {\bibfield
  {journal} {\bibinfo  {journal} {Rev. Mod. Phys.}\ }\textbf {\bibinfo {volume}
  {74}},\ \bibinfo {pages} {47} (\bibinfo {year} {2002})}\BibitemShut {NoStop}%
\bibitem [{\citenamefont {Newman}(2003)}]{Newman:2003p839}%
  \BibitemOpen
  \bibfield  {author} {\bibinfo {author} {\bibfnamefont {M.}~\bibnamefont
  {Newman}},\ }\href {\doibase 10.1137/S003614450342480} {\bibfield  {journal}
  {\bibinfo  {journal} {SIAM Rev.}\ }\textbf {\bibinfo {volume} {45}},\
  \bibinfo {pages} {167} (\bibinfo {year} {2003})}\BibitemShut {NoStop}%
\bibitem [{\citenamefont {Barrat}\ \emph {et~al.}(2008)\citenamefont {Barrat},
  \citenamefont {Barth{\'{e}}lemy},\ and\ \citenamefont
  {Vespignani}}]{Barrat2008}%
  \BibitemOpen
  \bibfield  {author} {\bibinfo {author} {\bibfnamefont {A.}~\bibnamefont
  {Barrat}}, \bibinfo {author} {\bibfnamefont {M.}~\bibnamefont
  {Barth{\'{e}}lemy}}, \ and\ \bibinfo {author} {\bibfnamefont
  {A.}~\bibnamefont {Vespignani}},\ }\href@noop {} {\emph {\bibinfo {title}
  {{Dynamical Processes on Complex Networks}}}}\ (\bibinfo  {publisher}
  {Cambridge: Cambridge University Press},\ \bibinfo {year} {2008})\BibitemShut
  {NoStop}%
\bibitem [{\citenamefont {Pastor-Satorras}\ \emph {et~al.}(2015)\citenamefont
  {Pastor-Satorras}, \citenamefont {Castellano}, \citenamefont {{Van
  Mieghem}},\ and\ \citenamefont {Vespignani}}]{Pastor-Satorras2015}%
  \BibitemOpen
  \bibfield  {author} {\bibinfo {author} {\bibfnamefont {R.}~\bibnamefont
  {Pastor-Satorras}}, \bibinfo {author} {\bibfnamefont {C.}~\bibnamefont
  {Castellano}}, \bibinfo {author} {\bibfnamefont {P.}~\bibnamefont {{Van
  Mieghem}}}, \ and\ \bibinfo {author} {\bibfnamefont {A.}~\bibnamefont
  {Vespignani}},\ }\href@noop {} {\bibfield  {journal} {\bibinfo  {journal}
  {Rev. Mod. Phys.}\ }\textbf {\bibinfo {volume} {87}},\ \bibinfo {pages} {925}
  (\bibinfo {year} {2015})}\BibitemShut {NoStop}%
\bibitem [{\citenamefont {Holme}\ and\ \citenamefont
  {Saram{\"a}ki}(2012)}]{Holme2012}%
  \BibitemOpen
  \bibfield  {author} {\bibinfo {author} {\bibfnamefont {P.}~\bibnamefont
  {Holme}}\ and\ \bibinfo {author} {\bibfnamefont {J.}~\bibnamefont
  {Saram{\"a}ki}},\ }\href {\doibase
  http://dx.doi.org/10.1016/j.physrep.2012.03.001} {\bibfield  {journal}
  {\bibinfo  {journal} {Phys. Rep.}\ }\textbf {\bibinfo {volume} {519}},\
  \bibinfo {pages} {97 } (\bibinfo {year} {2012})}\BibitemShut {NoStop}%
\bibitem [{\citenamefont {{Holme}}(2015)}]{Holme2015}%
  \BibitemOpen
  \bibfield  {author} {\bibinfo {author} {\bibfnamefont {P.}~\bibnamefont
  {{Holme}}},\ }\href {\doibase 10.1140/epjb/e2015-60657-4} {\bibfield
  {journal} {\bibinfo  {journal} {EPJ B}\ }\textbf {\bibinfo {volume} {88}},\
  \bibinfo {eid} {234} (\bibinfo {year} {2015})}\BibitemShut {NoStop}%
\bibitem [{\citenamefont {Masada}\ and\ \citenamefont
  {Lambiotte}(2016)}]{MasadaLambiotte2016}%
  \BibitemOpen
  \bibfield  {author} {\bibinfo {author} {\bibfnamefont {N.}~\bibnamefont
  {Masada}}\ and\ \bibinfo {author} {\bibfnamefont {R.}~\bibnamefont
  {Lambiotte}},\ }\href@noop {} {\emph {\bibinfo {title} {A guide to temporal
  networks}}}\ (\bibinfo  {publisher} {World Scientific},\ \bibinfo {year}
  {2016})\BibitemShut {NoStop}%
\bibitem [{\citenamefont {Sanli}\ and\ \citenamefont
  {Lambiotte}(2015{\natexlab{a}})}]{Sanli2015}%
  \BibitemOpen
  \bibfield  {author} {\bibinfo {author} {\bibfnamefont {C.}~\bibnamefont
  {Sanli}}\ and\ \bibinfo {author} {\bibfnamefont {R.}~\bibnamefont
  {Lambiotte}},\ }\href@noop {} {\bibfield  {journal} {\bibinfo  {journal}
  {Frontiers in Physics}\ }\textbf {\bibinfo {volume} {3}},\ \bibinfo {pages}
  {79} (\bibinfo {year} {2015}{\natexlab{a}})}\BibitemShut {NoStop}%
\bibitem [{\citenamefont {Borge-Holthoefer}\ \emph {et~al.}(2016)\citenamefont
  {Borge-Holthoefer}, \citenamefont {Perra}, \citenamefont {Gon{\c c}alves},
  \citenamefont {Gonz{\'a}lez-Bail{\'o}n}, \citenamefont {Arenas},
  \citenamefont {Moreno},\ and\ \citenamefont
  {Vespignani}}]{Borge-Holthoefer2015}%
  \BibitemOpen
  \bibfield  {author} {\bibinfo {author} {\bibfnamefont {J.}~\bibnamefont
  {Borge-Holthoefer}}, \bibinfo {author} {\bibfnamefont {N.}~\bibnamefont
  {Perra}}, \bibinfo {author} {\bibfnamefont {B.}~\bibnamefont {Gon{\c
  c}alves}}, \bibinfo {author} {\bibfnamefont {S.}~\bibnamefont
  {Gonz{\'a}lez-Bail{\'o}n}}, \bibinfo {author} {\bibfnamefont
  {A.}~\bibnamefont {Arenas}}, \bibinfo {author} {\bibfnamefont
  {Y.}~\bibnamefont {Moreno}}, \ and\ \bibinfo {author} {\bibfnamefont
  {A.}~\bibnamefont {Vespignani}},\ }\href@noop {} {\bibfield  {journal}
  {\bibinfo  {journal} {Science Advances}\ }\textbf {\bibinfo {volume} {2}},\
  \bibinfo {pages} {e1501158} (\bibinfo {year} {2016})}\BibitemShut {NoStop}%
\bibitem [{\citenamefont {Darst}\ \emph {et~al.}(2016)\citenamefont {Darst},
  \citenamefont {Granell}, \citenamefont {Arenas}, \citenamefont {G{\'{o}}mez},
  \citenamefont {Saram{\"{a}}ki},\ and\ \citenamefont {Fortunato}}]{Darst2016}%
  \BibitemOpen
  \bibfield  {author} {\bibinfo {author} {\bibfnamefont {R.~K.}\ \bibnamefont
  {Darst}}, \bibinfo {author} {\bibfnamefont {C.}~\bibnamefont {Granell}},
  \bibinfo {author} {\bibfnamefont {A.}~\bibnamefont {Arenas}}, \bibinfo
  {author} {\bibfnamefont {S.}~\bibnamefont {G{\'{o}}mez}}, \bibinfo {author}
  {\bibfnamefont {J.}~\bibnamefont {Saram{\"{a}}ki}}, \ and\ \bibinfo {author}
  {\bibfnamefont {S.}~\bibnamefont {Fortunato}},\ }\href@noop {} {\enquote
  {\bibinfo {title} {{Detection of timescales in evolving complex systems}},}\
  } (\bibinfo {year} {2016}),\ \Eprint {http://arxiv.org/abs/1604.00758}
  {arXiv:1604.00758} \BibitemShut {NoStop}%
\bibitem [{\citenamefont {Eagle}\ and\ \citenamefont {{(Sandy)
  Pentland}}(2005)}]{Eagle2005}%
  \BibitemOpen
  \bibfield  {author} {\bibinfo {author} {\bibfnamefont {N.}~\bibnamefont
  {Eagle}}\ and\ \bibinfo {author} {\bibfnamefont {A.}~\bibnamefont {{(Sandy)
  Pentland}}},\ }\href@noop {} {\bibfield  {journal} {\bibinfo  {journal}
  {Pers. Ubiquit. Comput.}\ }\textbf {\bibinfo {volume} {10}},\ \bibinfo
  {pages} {255} (\bibinfo {year} {2005})}\BibitemShut {NoStop}%
\bibitem [{\citenamefont {Isella}\ \emph {et~al.}(2011)\citenamefont {Isella},
  \citenamefont {Romano}, \citenamefont {Barrat}, \citenamefont {Cattuto},
  \citenamefont {Colizza}, \citenamefont {{Van den Broeck}}, \citenamefont
  {Gesualdo}, \citenamefont {Pandolfi}, \citenamefont {Rav{\`{a}}},
  \citenamefont {Rizzo},\ and\ \citenamefont {Tozzi}}]{Isella2011}%
  \BibitemOpen
  \bibfield  {author} {\bibinfo {author} {\bibfnamefont {L.}~\bibnamefont
  {Isella}}, \bibinfo {author} {\bibfnamefont {M.}~\bibnamefont {Romano}},
  \bibinfo {author} {\bibfnamefont {A.}~\bibnamefont {Barrat}}, \bibinfo
  {author} {\bibfnamefont {C.}~\bibnamefont {Cattuto}}, \bibinfo {author}
  {\bibfnamefont {V.}~\bibnamefont {Colizza}}, \bibinfo {author} {\bibfnamefont
  {W.}~\bibnamefont {{Van den Broeck}}}, \bibinfo {author} {\bibfnamefont
  {F.}~\bibnamefont {Gesualdo}}, \bibinfo {author} {\bibfnamefont
  {E.}~\bibnamefont {Pandolfi}}, \bibinfo {author} {\bibfnamefont
  {L.}~\bibnamefont {Rav{\`{a}}}}, \bibinfo {author} {\bibfnamefont
  {C.}~\bibnamefont {Rizzo}}, \ and\ \bibinfo {author} {\bibfnamefont {A.~E.}\
  \bibnamefont {Tozzi}},\ }\href@noop {} {\bibfield  {journal} {\bibinfo
  {journal} {PLoS ONE}\ }\textbf {\bibinfo {volume} {6}},\ \bibinfo {pages}
  {e17144} (\bibinfo {year} {2011})}\BibitemShut {NoStop}%
\bibitem [{\citenamefont {Eckmann}\ \emph {et~al.}(2004)\citenamefont
  {Eckmann}, \citenamefont {Moses},\ and\ \citenamefont {Sergi}}]{Eckmann2004}%
  \BibitemOpen
  \bibfield  {author} {\bibinfo {author} {\bibfnamefont {J.~P.}\ \bibnamefont
  {Eckmann}}, \bibinfo {author} {\bibfnamefont {E.}~\bibnamefont {Moses}}, \
  and\ \bibinfo {author} {\bibfnamefont {D.}~\bibnamefont {Sergi}},\
  }\href@noop {} {\bibfield  {journal} {\bibinfo  {journal} {Proc. Natl. Acad.
  Sci. USA}\ }\textbf {\bibinfo {volume} {101}},\ \bibinfo {pages} {14333}
  (\bibinfo {year} {2004})}\BibitemShut {NoStop}%
\bibitem [{\citenamefont {Barabasi}(2005)}]{Barabasi:2005aa}%
  \BibitemOpen
  \bibfield  {author} {\bibinfo {author} {\bibfnamefont {A.-L.}\ \bibnamefont
  {Barabasi}},\ }\href@noop {} {\bibfield  {journal} {\bibinfo  {journal}
  {Nature}\ }\textbf {\bibinfo {volume} {435}},\ \bibinfo {pages} {207}
  (\bibinfo {year} {2005})}\BibitemShut {NoStop}%
\bibitem [{\citenamefont {V\'azquez}\ \emph {et~al.}(2006)\citenamefont
  {V\'azquez}, \citenamefont {Oliveira}, \citenamefont {Dezs\"o}, \citenamefont
  {Goh}, \citenamefont {Kondor},\ and\ \citenamefont
  {Barab\'asi}}]{Alexei2006PRE}%
  \BibitemOpen
  \bibfield  {author} {\bibinfo {author} {\bibfnamefont {A.}~\bibnamefont
  {V\'azquez}}, \bibinfo {author} {\bibfnamefont {J.~G.}\ \bibnamefont
  {Oliveira}}, \bibinfo {author} {\bibfnamefont {Z.}~\bibnamefont {Dezs\"o}},
  \bibinfo {author} {\bibfnamefont {K.-I.}\ \bibnamefont {Goh}}, \bibinfo
  {author} {\bibfnamefont {I.}~\bibnamefont {Kondor}}, \ and\ \bibinfo {author}
  {\bibfnamefont {A.-L.}\ \bibnamefont {Barab\'asi}},\ }\href {\doibase
  10.1103/PhysRevE.73.036127} {\bibfield  {journal} {\bibinfo  {journal} {Phys.
  Rev. E}\ }\textbf {\bibinfo {volume} {73}},\ \bibinfo {pages} {036127}
  (\bibinfo {year} {2006})}\BibitemShut {NoStop}%
\bibitem [{\citenamefont {Malmgren}\ \emph {et~al.}(2008)\citenamefont
  {Malmgren}, \citenamefont {Stouffer}, \citenamefont {Motter},\ and\
  \citenamefont {Amaral}}]{Malmgren2008}%
  \BibitemOpen
  \bibfield  {author} {\bibinfo {author} {\bibfnamefont {R.}~\bibnamefont
  {Malmgren}}, \bibinfo {author} {\bibfnamefont {D.}~\bibnamefont {Stouffer}},
  \bibinfo {author} {\bibfnamefont {A.}~\bibnamefont {Motter}}, \ and\ \bibinfo
  {author} {\bibfnamefont {L.}~\bibnamefont {Amaral}},\ }\href@noop {}
  {\bibfield  {journal} {\bibinfo  {journal} {Proc. Natl. Acad. Sci. USA}\
  }\textbf {\bibinfo {volume} {105}},\ \bibinfo {pages} {18153} (\bibinfo
  {year} {2008})}\BibitemShut {NoStop}%
\bibitem [{\citenamefont {Goh}\ and\ \citenamefont
  {Barab{\'a}si}(2008)}]{Goh2008}%
  \BibitemOpen
  \bibfield  {author} {\bibinfo {author} {\bibfnamefont {K.-I.}\ \bibnamefont
  {Goh}}\ and\ \bibinfo {author} {\bibfnamefont {A.-L.}\ \bibnamefont
  {Barab{\'a}si}},\ }\href@noop {} {\bibfield  {journal} {\bibinfo  {journal}
  {Europhys. Lett.}\ }\textbf {\bibinfo {volume} {81}},\ \bibinfo {pages}
  {48002} (\bibinfo {year} {2008})}\BibitemShut {NoStop}%
\bibitem [{\citenamefont {Karsai}\ \emph {et~al.}(2012)\citenamefont {Karsai},
  \citenamefont {Kaski}, \citenamefont {Barab{\'a}si},\ and\ \citenamefont
  {Kert{\'e}sz}}]{Karsai2012}%
  \BibitemOpen
  \bibfield  {author} {\bibinfo {author} {\bibfnamefont {M.}~\bibnamefont
  {Karsai}}, \bibinfo {author} {\bibfnamefont {K.}~\bibnamefont {Kaski}},
  \bibinfo {author} {\bibfnamefont {A.-L.}\ \bibnamefont {Barab{\'a}si}}, \
  and\ \bibinfo {author} {\bibfnamefont {J.}~\bibnamefont {Kert{\'e}sz}},\
  }\href@noop {} {\bibfield  {journal} {\bibinfo  {journal} {Sci. Rep.}\
  }\textbf {\bibinfo {volume} {2}},\ \bibinfo {pages} {397} (\bibinfo {year}
  {2012})}\BibitemShut {NoStop}%
\bibitem [{\citenamefont {Vestergaard}\ \emph {et~al.}(2014)\citenamefont
  {Vestergaard}, \citenamefont {G{\'{e}}nois},\ and\ \citenamefont
  {Barrat}}]{Vestergaard2014}%
  \BibitemOpen
  \bibfield  {author} {\bibinfo {author} {\bibfnamefont {C.~L.}\ \bibnamefont
  {Vestergaard}}, \bibinfo {author} {\bibfnamefont {M.}~\bibnamefont
  {G{\'{e}}nois}}, \ and\ \bibinfo {author} {\bibfnamefont {A.}~\bibnamefont
  {Barrat}},\ }\href@noop {} {\bibfield  {journal} {\bibinfo  {journal} {Phys.
  Rev. E}\ }\textbf {\bibinfo {volume} {90}},\ \bibinfo {pages} {042805}
  (\bibinfo {year} {2014})}\BibitemShut {NoStop}%
\bibitem [{\citenamefont {Dayan}\ and\ \citenamefont
  {Abbott}(2001)}]{dayan_abbott_2001tnc}%
  \BibitemOpen
  \bibfield  {author} {\bibinfo {author} {\bibfnamefont {P.}~\bibnamefont
  {Dayan}}\ and\ \bibinfo {author} {\bibfnamefont {L.~F.}\ \bibnamefont
  {Abbott}},\ }\href@noop {} {\emph {\bibinfo {title} {Theoretical
  Neuroscience: Computational and Mathematical Modeling of Neural Systems}}}\
  (\bibinfo  {publisher} {MIT Press},\ \bibinfo {year} {2001})\BibitemShut
  {NoStop}%
\bibitem [{\citenamefont {Gabbiani}\ and\ \citenamefont
  {Koch}(1998)}]{gabbiani1998}%
  \BibitemOpen
  \bibfield  {author} {\bibinfo {author} {\bibfnamefont {F.}~\bibnamefont
  {Gabbiani}}\ and\ \bibinfo {author} {\bibfnamefont {C.}~\bibnamefont
  {Koch}},\ }in\ \href@noop {} {\emph {\bibinfo {booktitle} {Methods of
  Neuronal Modeling}}},\ \bibinfo {editor} {edited by\ \bibinfo {editor}
  {\bibfnamefont {C.}~\bibnamefont {Koch}}\ and\ \bibinfo {editor}
  {\bibfnamefont {I.}~\bibnamefont {Segev}}}\ (\bibinfo  {publisher} {MIT
  Press},\ \bibinfo {year} {1998})\ pp.\ \bibinfo {pages}
  {313--360}\BibitemShut {NoStop}%
\bibitem [{\citenamefont {Kostal}\ \emph {et~al.}(2007)\citenamefont {Kostal},
  \citenamefont {Lansky},\ and\ \citenamefont {Rospars}}]{kostal2007}%
  \BibitemOpen
  \bibfield  {author} {\bibinfo {author} {\bibfnamefont {L.}~\bibnamefont
  {Kostal}}, \bibinfo {author} {\bibfnamefont {P.}~\bibnamefont {Lansky}}, \
  and\ \bibinfo {author} {\bibfnamefont {J.-P.}\ \bibnamefont {Rospars}},\
  }\href@noop {} {\bibfield  {journal} {\bibinfo  {journal} {Eur. J.
  Neurosci.}\ }\textbf {\bibinfo {volume} {26}},\ \bibinfo {pages} {2693}
  (\bibinfo {year} {2007})}\BibitemShut {NoStop}%
\bibitem [{\citenamefont {Shinomoto}\ \emph {et~al.}(2003)\citenamefont
  {Shinomoto}, \citenamefont {Shima},\ and\ \citenamefont
  {Tanji}}]{Shinomoto:2003aa}%
  \BibitemOpen
  \bibfield  {author} {\bibinfo {author} {\bibfnamefont {S.}~\bibnamefont
  {Shinomoto}}, \bibinfo {author} {\bibfnamefont {K.}~\bibnamefont {Shima}}, \
  and\ \bibinfo {author} {\bibfnamefont {J.}~\bibnamefont {Tanji}},\ }\href
  {\doibase 10.1162/089976603322518759} {\bibfield  {journal} {\bibinfo
  {journal} {Neural Comput.}\ }\textbf {\bibinfo {volume} {15}},\ \bibinfo
  {pages} {2823} (\bibinfo {year} {2003})}\BibitemShut {NoStop}%
\bibitem [{\citenamefont {Shinomoto}\ \emph {et~al.}(2009)\citenamefont
  {Shinomoto}, \citenamefont {Kim}, \citenamefont {Shimokawa}, \citenamefont
  {Matsuno}, \citenamefont {Funahashi}, \citenamefont {Shima}, \citenamefont
  {Fujita}, \citenamefont {Tamura}, \citenamefont {Doi}, \citenamefont
  {Kawano}, \citenamefont {Inaba}, \citenamefont {Fukushima}, \citenamefont
  {Kurkin}, \citenamefont {Kurata}, \citenamefont {Taira}, \citenamefont
  {Tsutsui}, \citenamefont {Komatsu}, \citenamefont {Ogawa}, \citenamefont
  {Koida}, \citenamefont {Tanji},\ and\ \citenamefont
  {Toyama}}]{Shinomoto2009}%
  \BibitemOpen
  \bibfield  {author} {\bibinfo {author} {\bibfnamefont {S.}~\bibnamefont
  {Shinomoto}}, \bibinfo {author} {\bibfnamefont {H.}~\bibnamefont {Kim}},
  \bibinfo {author} {\bibfnamefont {T.}~\bibnamefont {Shimokawa}}, \bibinfo
  {author} {\bibfnamefont {N.}~\bibnamefont {Matsuno}}, \bibinfo {author}
  {\bibfnamefont {S.}~\bibnamefont {Funahashi}}, \bibinfo {author}
  {\bibfnamefont {K.}~\bibnamefont {Shima}}, \bibinfo {author} {\bibfnamefont
  {I.}~\bibnamefont {Fujita}}, \bibinfo {author} {\bibfnamefont
  {H.}~\bibnamefont {Tamura}}, \bibinfo {author} {\bibfnamefont
  {T.}~\bibnamefont {Doi}}, \bibinfo {author} {\bibfnamefont {K.}~\bibnamefont
  {Kawano}}, \bibinfo {author} {\bibfnamefont {N.}~\bibnamefont {Inaba}},
  \bibinfo {author} {\bibfnamefont {K.}~\bibnamefont {Fukushima}}, \bibinfo
  {author} {\bibfnamefont {S.}~\bibnamefont {Kurkin}}, \bibinfo {author}
  {\bibfnamefont {K.}~\bibnamefont {Kurata}}, \bibinfo {author} {\bibfnamefont
  {M.}~\bibnamefont {Taira}}, \bibinfo {author} {\bibfnamefont {K.-I.}\
  \bibnamefont {Tsutsui}}, \bibinfo {author} {\bibfnamefont {H.}~\bibnamefont
  {Komatsu}}, \bibinfo {author} {\bibfnamefont {T.}~\bibnamefont {Ogawa}},
  \bibinfo {author} {\bibfnamefont {K.}~\bibnamefont {Koida}}, \bibinfo
  {author} {\bibfnamefont {J.}~\bibnamefont {Tanji}}, \ and\ \bibinfo {author}
  {\bibfnamefont {K.}~\bibnamefont {Toyama}},\ }\href {\doibase
  10.1371/journal.pcbi.1000433} {\bibfield  {journal} {\bibinfo  {journal}
  {PLoS Comput. Biol.}\ }\textbf {\bibinfo {volume} {5}},\ \bibinfo {pages}
  {e1000433} (\bibinfo {year} {2009})}\BibitemShut {NoStop}%
\bibitem [{\citenamefont {Shinomoto}\ \emph {et~al.}(2005)\citenamefont
  {Shinomoto}, \citenamefont {Miura},\ and\ \citenamefont
  {Koyama}}]{shinomoto2005}%
  \BibitemOpen
  \bibfield  {author} {\bibinfo {author} {\bibfnamefont {S.}~\bibnamefont
  {Shinomoto}}, \bibinfo {author} {\bibfnamefont {K.}~\bibnamefont {Miura}}, \
  and\ \bibinfo {author} {\bibfnamefont {S.}~\bibnamefont {Koyama}},\
  }\href@noop {} {\bibfield  {journal} {\bibinfo  {journal} {Biosystems}\
  }\textbf {\bibinfo {volume} {79}},\ \bibinfo {pages} {67} (\bibinfo {year}
  {2005})}\BibitemShut {NoStop}%
\bibitem [{\citenamefont {Miura}\ \emph {et~al.}(2006)\citenamefont {Miura},
  \citenamefont {Okada},\ and\ \citenamefont {Amari}}]{Miura:2006aa}%
  \BibitemOpen
  \bibfield  {author} {\bibinfo {author} {\bibfnamefont {K.}~\bibnamefont
  {Miura}}, \bibinfo {author} {\bibfnamefont {M.}~\bibnamefont {Okada}}, \ and\
  \bibinfo {author} {\bibfnamefont {S.-i.}\ \bibnamefont {Amari}},\ }\bibfield
  {booktitle} {\emph {\bibinfo {booktitle} {Neural Computation}},\ }\href
  {\doibase 10.1162/neco.2006.18.10.2359} {\bibfield  {journal} {\bibinfo
  {journal} {Neural Comput.}\ }\textbf {\bibinfo {volume} {18}},\ \bibinfo
  {pages} {2359} (\bibinfo {year} {2006})}\BibitemShut {NoStop}%
\bibitem [{\citenamefont {Jo}\ \emph {et~al.}(2012)\citenamefont {Jo},
  \citenamefont {Karsai}, \citenamefont {Kert{\'e}sz},\ and\ \citenamefont
  {Kaski}}]{Jo2012}%
  \BibitemOpen
  \bibfield  {author} {\bibinfo {author} {\bibfnamefont {H.-H.}\ \bibnamefont
  {Jo}}, \bibinfo {author} {\bibfnamefont {M.}~\bibnamefont {Karsai}}, \bibinfo
  {author} {\bibfnamefont {J.}~\bibnamefont {Kert{\'e}sz}}, \ and\ \bibinfo
  {author} {\bibfnamefont {K.}~\bibnamefont {Kaski}},\ }\href
  {http://stacks.iop.org/1367-2630/14/i=1/a=013055} {\bibfield  {journal}
  {\bibinfo  {journal} {New J. Phys.}\ }\textbf {\bibinfo {volume} {14}},\
  \bibinfo {pages} {013055} (\bibinfo {year} {2012})}\BibitemShut {NoStop}%
\bibitem [{\citenamefont {Kobayashi}\ and\ \citenamefont
  {Lambiotte}(2016)}]{KobaLambiotte2016}%
  \BibitemOpen
  \bibfield  {author} {\bibinfo {author} {\bibfnamefont {R.}~\bibnamefont
  {Kobayashi}}\ and\ \bibinfo {author} {\bibfnamefont {R.}~\bibnamefont
  {Lambiotte}},\ }in\ \href@noop {} {\emph {\bibinfo {booktitle} {Proceedings
  of the 10th International AAAI Conference on Web and Social Media, Cologne,
  2016}}}\ (\bibinfo  {publisher} {The AAI Press},\ \bibinfo {address} {Plto,
  CA, USA},\ \bibinfo {year} {2016})\ p.\ \bibinfo {pages} {191}\BibitemShut
  {NoStop}%
\bibitem [{\citenamefont {Sanli}\ and\ \citenamefont
  {Lambiotte}(2015{\natexlab{b}})}]{CeydaRenaud}%
  \BibitemOpen
  \bibfield  {author} {\bibinfo {author} {\bibfnamefont {C.}~\bibnamefont
  {Sanli}}\ and\ \bibinfo {author} {\bibfnamefont {R.}~\bibnamefont
  {Lambiotte}},\ }\href {\doibase 10.1371/journal.pone.0131704} {\bibfield
  {journal} {\bibinfo  {journal} {PLoS ONE}\ }\textbf {\bibinfo {volume}
  {10}},\ \bibinfo {pages} {e0131704} (\bibinfo {year}
  {2015}{\natexlab{b}})}\BibitemShut {NoStop}%
\bibitem [{\citenamefont {Tabourier}\ \emph {et~al.}(2016)\citenamefont
  {Tabourier}, \citenamefont {Libert},\ and\ \citenamefont
  {Lambiotte}}]{Tabourier:2016ks}%
  \BibitemOpen
  \bibfield  {author} {\bibinfo {author} {\bibfnamefont {L.}~\bibnamefont
  {Tabourier}}, \bibinfo {author} {\bibfnamefont {A.-S.}\ \bibnamefont
  {Libert}}, \ and\ \bibinfo {author} {\bibfnamefont {R.}~\bibnamefont
  {Lambiotte}},\ }\href@noop {} {\bibfield  {journal} {\bibinfo  {journal} {EPJ
  Data Science}\ }\textbf {\bibinfo {volume} {5}},\ \bibinfo {pages} {1}
  (\bibinfo {year} {2016})}\BibitemShut {NoStop}%
\bibitem [{\citenamefont {Klimt}\ and\ \citenamefont
  {Yang}(2004)}]{konect:klimt04}%
  \BibitemOpen
  \bibfield  {author} {\bibinfo {author} {\bibfnamefont {B.}~\bibnamefont
  {Klimt}}\ and\ \bibinfo {author} {\bibfnamefont {Y.}~\bibnamefont {Yang}},\
  }in\ \href@noop {} {\emph {\bibinfo {booktitle} {Proceedings of 15th European
  Conference on Machine Learning, Pisa, Italy, 2004}}}\ (\bibinfo  {publisher}
  {Springer},\ \bibinfo {address} {Berlin Heidelberg},\ \bibinfo {year}
  {2004})\ pp.\ \bibinfo {pages} {217--226}\BibitemShut {NoStop}%
\bibitem [{Note1()}]{Note1}%
  \BibitemOpen
  \bibinfo {note} {Http://konect.uni-koblenz.de/ (Date accessed: March 22nd,
  2016).}\BibitemShut {Stop}%
\bibitem [{\citenamefont {Shepherd}(2004)}]{shepherd2004synaptic}%
  \BibitemOpen
  \bibfield  {author} {\bibinfo {author} {\bibfnamefont {G.}~\bibnamefont
  {Shepherd}},\ }\href {https://books.google.co.jp/books?id=xlCqS0EODUoC}
  {\emph {\bibinfo {title} {The Synaptic Organization of the Brain}}}\
  (\bibinfo  {publisher} {Oxford University Press, USA},\ \bibinfo {year}
  {2004})\BibitemShut {NoStop}%
\bibitem [{\citenamefont {Ahtola}(1978)}]{Wildt1978}%
  \BibitemOpen
  \bibfield  {author} {\bibinfo {author} {\bibfnamefont {A.~R. W.~O.}\
  \bibnamefont {Ahtola}},\ }\href@noop {} {\emph {\bibinfo {title} {Analysis of
  Covariance}}}\ (\bibinfo  {publisher} {Thousand Oaks: SAGE Publications,
  Inc.},\ \bibinfo {year} {1978})\BibitemShut {NoStop}%
\bibitem [{\citenamefont {Kass}\ \emph {et~al.}(2005)\citenamefont {Kass},
  \citenamefont {Ventura},\ and\ \citenamefont {Brown}}]{Kass:2005go}%
  \BibitemOpen
  \bibfield  {author} {\bibinfo {author} {\bibfnamefont {R.~E.}\ \bibnamefont
  {Kass}}, \bibinfo {author} {\bibfnamefont {V.}~\bibnamefont {Ventura}}, \
  and\ \bibinfo {author} {\bibfnamefont {E.~N.}\ \bibnamefont {Brown}},\
  }\href@noop {} {\bibfield  {journal} {\bibinfo  {journal} {J. Neurophysiol.}\
  }\textbf {\bibinfo {volume} {94}},\ \bibinfo {pages} {8} (\bibinfo {year}
  {2005})}\BibitemShut {NoStop}%
\bibitem [{\citenamefont {Backlund}\ \emph {et~al.}(2014)\citenamefont
  {Backlund}, \citenamefont {Saram\"aki},\ and\ \citenamefont
  {Pan}}]{Saramaki2015PRE}%
  \BibitemOpen
  \bibfield  {author} {\bibinfo {author} {\bibfnamefont {V.-P.}\ \bibnamefont
  {Backlund}}, \bibinfo {author} {\bibfnamefont {J.}~\bibnamefont
  {Saram\"aki}}, \ and\ \bibinfo {author} {\bibfnamefont {R.~K.}\ \bibnamefont
  {Pan}},\ }\href@noop {} {\bibfield  {journal} {\bibinfo  {journal} {Phys.
  Rev. E}\ }\textbf {\bibinfo {volume} {89}},\ \bibinfo {pages} {062815}
  (\bibinfo {year} {2014})}\BibitemShut {NoStop}%
\bibitem [{\citenamefont {Holme}(2005)}]{PhysRevE.71.046119}%
  \BibitemOpen
  \bibfield  {author} {\bibinfo {author} {\bibfnamefont {P.}~\bibnamefont
  {Holme}},\ }\href {\doibase 10.1103/PhysRevE.71.046119} {\bibfield  {journal}
  {\bibinfo  {journal} {Phys. Rev. E}\ }\textbf {\bibinfo {volume} {71}},\
  \bibinfo {pages} {046119} (\bibinfo {year} {2005})}\BibitemShut {NoStop}%
\bibitem [{\citenamefont {Hawkes}(1971)}]{Hawkes}%
  \BibitemOpen
  \bibfield  {author} {\bibinfo {author} {\bibfnamefont {A.~G.}\ \bibnamefont
  {Hawkes}},\ }\href {http://www.jstor.org/stable/2334319} {\bibfield
  {journal} {\bibinfo  {journal} {Biometrika}\ }\textbf {\bibinfo {volume}
  {58}},\ \bibinfo {pages} {83} (\bibinfo {year} {1971})}\BibitemShut {NoStop}%
\bibitem [{\citenamefont {Ogata}(1988)}]{ogata1988}%
  \BibitemOpen
  \bibfield  {author} {\bibinfo {author} {\bibfnamefont {Y.}~\bibnamefont
  {Ogata}},\ }\href@noop {} {\bibfield  {journal} {\bibinfo  {journal} {J.
  Amer. Statist. Assoc.}\ }\textbf {\bibinfo {volume} {83}},\ \bibinfo {pages}
  {9} (\bibinfo {year} {1988})}\BibitemShut {NoStop}%
\bibitem [{\citenamefont {Masuda}\ \emph {et~al.}(2013)\citenamefont {Masuda},
  \citenamefont {Takaguchi}, \citenamefont {Sato},\ and\ \citenamefont
  {Yano}}]{masuda2013}%
  \BibitemOpen
  \bibfield  {author} {\bibinfo {author} {\bibfnamefont {N.}~\bibnamefont
  {Masuda}}, \bibinfo {author} {\bibfnamefont {T.}~\bibnamefont {Takaguchi}},
  \bibinfo {author} {\bibfnamefont {N.}~\bibnamefont {Sato}}, \ and\ \bibinfo
  {author} {\bibfnamefont {K.}~\bibnamefont {Yano}},\ }in\ \href {\doibase
  10.1007/978-3-642-36461-7_12} {\emph {\bibinfo {booktitle} {Temporal
  Networks}}},\ \bibinfo {editor} {edited by\ \bibinfo {editor} {\bibfnamefont
  {P.}~\bibnamefont {Holme}}\ and\ \bibinfo {editor} {\bibfnamefont
  {J.}~\bibnamefont {Saram{\"a}ki}}}\ (\bibinfo  {publisher} {Springer Berlin
  Heidelberg},\ \bibinfo {year} {2013})\ pp.\ \bibinfo {pages}
  {245--264}\BibitemShut {NoStop}%
\bibitem [{\citenamefont {Onaga}\ and\ \citenamefont
  {Shinomoto}(2014)}]{PhysRevE.89.042817}%
  \BibitemOpen
  \bibfield  {author} {\bibinfo {author} {\bibfnamefont {T.}~\bibnamefont
  {Onaga}}\ and\ \bibinfo {author} {\bibfnamefont {S.}~\bibnamefont
  {Shinomoto}},\ }\href {\doibase 10.1103/PhysRevE.89.042817} {\bibfield
  {journal} {\bibinfo  {journal} {Phys. Rev. E}\ }\textbf {\bibinfo {volume}
  {89}},\ \bibinfo {pages} {042817} (\bibinfo {year} {2014})}\BibitemShut
  {NoStop}%
\bibitem [{\citenamefont {Zhao}\ \emph {et~al.}(2015)\citenamefont {Zhao},
  \citenamefont {Erdogdu}, \citenamefont {He}, \citenamefont {Rajaraman},\ and\
  \citenamefont {Leskovec}}]{SEISMIC}%
  \BibitemOpen
  \bibfield  {author} {\bibinfo {author} {\bibfnamefont {Q.}~\bibnamefont
  {Zhao}}, \bibinfo {author} {\bibfnamefont {M.~A.}\ \bibnamefont {Erdogdu}},
  \bibinfo {author} {\bibfnamefont {H.~Y.}\ \bibnamefont {He}}, \bibinfo
  {author} {\bibfnamefont {A.}~\bibnamefont {Rajaraman}}, \ and\ \bibinfo
  {author} {\bibfnamefont {J.}~\bibnamefont {Leskovec}},\ }in\ \href {\doibase
  10.1145/2783258.2783401} {\emph {\bibinfo {booktitle} {Proceedings of the
  21th ACM SIGKDD International Conference on Knowledge Discovery and Data
  Mining, New York, 2015}}}\ (\bibinfo  {publisher} {ACM},\ \bibinfo {address}
  {New York, NY, USA},\ \bibinfo {year} {2015})\ pp.\ \bibinfo {pages}
  {1513--1522}\BibitemShut {NoStop}%
\bibitem [{Note2()}]{Note2}%
  \BibitemOpen
  \bibinfo {note} {In the empirical data analysis, the sequence that has at
  least $100$ IETs are selected to evaluate LV and CV measures (see
  section~\ref {sec:LV_CV}).}\BibitemShut {Stop}%
\bibitem [{\citenamefont {Jolivet}\ \emph {et~al.}(2008)\citenamefont
  {Jolivet}, \citenamefont {Kobayashi}, \citenamefont {Rauch}, \citenamefont
  {Naud}, \citenamefont {Shinomoto},\ and\ \citenamefont
  {Gerstner}}]{jolivet2008}%
  \BibitemOpen
  \bibfield  {author} {\bibinfo {author} {\bibfnamefont {R.}~\bibnamefont
  {Jolivet}}, \bibinfo {author} {\bibfnamefont {R.}~\bibnamefont {Kobayashi}},
  \bibinfo {author} {\bibfnamefont {A.}~\bibnamefont {Rauch}}, \bibinfo
  {author} {\bibfnamefont {R.}~\bibnamefont {Naud}}, \bibinfo {author}
  {\bibfnamefont {S.}~\bibnamefont {Shinomoto}}, \ and\ \bibinfo {author}
  {\bibfnamefont {W.}~\bibnamefont {Gerstner}},\ }\href@noop {} {\bibfield
  {journal} {\bibinfo  {journal} {J. Neurosci. Methods}\ }\textbf {\bibinfo
  {volume} {169}},\ \bibinfo {pages} {417} (\bibinfo {year}
  {2008})}\BibitemShut {NoStop}%
\bibitem [{\citenamefont {Kobayashi}\ \emph {et~al.}(2009)\citenamefont
  {Kobayashi}, \citenamefont {Tsubo},\ and\ \citenamefont
  {Shinomoto}}]{MATMODEL}%
  \BibitemOpen
  \bibfield  {author} {\bibinfo {author} {\bibfnamefont {R.}~\bibnamefont
  {Kobayashi}}, \bibinfo {author} {\bibfnamefont {Y.}~\bibnamefont {Tsubo}}, \
  and\ \bibinfo {author} {\bibfnamefont {S.}~\bibnamefont {Shinomoto}},\
  }\href@noop {} {\bibfield  {journal} {\bibinfo  {journal} {Front. Comput.
  Neurosci.}\ }\textbf {\bibinfo {volume} {3}},\ \bibinfo {pages} {9} (\bibinfo
  {year} {2009})}\BibitemShut {NoStop}%
\end{thebibliography}

\end{document}